\documentclass[sigconf,nonacm]{acmart}

\makeatletter
\newcommand{\myconfshort}{\acmConference@shortname}
\newcommand{\myconffull}{\acmConference@name}
\newcommand{\myconfdate}{\acmConference@date}
\newcommand{\myconfloc}{\acmConference@venue}
\AtBeginDocument{
  \fancypagestyle{firstpagestyle}{
    \fancyhead{}%
    \fancyfoot[C]{}%
  }
  \fancyhf{}
  \fancyhead[LO]{\@headfootfont\shorttitle}%
  \fancyhead[RE]{\@headfootfont\@shortauthors}%
  \fancyhead[LE]{\@headfootfont\footnotesize \myconfshort, \myconfdate, \myconfloc}%
  \fancyhead[RO]{\@headfootfont\footnotesize \myconfshort, \myconfdate, \myconfloc}%
  \fancyfoot[C]{}%
}
\makeatother
\acmBooktitle{\conffull\@ (\confshort), \confdate, \confloc}

\AtBeginDocument{%
  }

\copyrightyear{2026}
\acmYear{2026}
\setcopyright{cc}
\setcctype{by}
\acmConference[FAccT '26]{The 2026 ACM Conference on Fairness, Accountability, and Transparency}{June 25--28, 2026}{Montreal, QC, Canada}
\acmBooktitle{The 2026 ACM Conference on Fairness, Accountability, and Transparency (FAccT '26), June 25--28, 2026, Montreal, QC, Canada}
\acmDOI{10.1145/3805689.3812218}
\acmISBN{979-8-4007-2596-8/2026/06}

\usepackage{threeparttable}
\usepackage{booktabs}
\usepackage{xcolor}  
\usepackage{colortbl}  
\usepackage{subcaption}
\usepackage{wrapfig}
\usepackage{float}
\usepackage{graphicx}
\usepackage{balance}
\usepackage{float}
\usepackage{tcolorbox}
\tcbuselibrary{skins}
\usepackage{afterpage}
\usepackage{booktabs}
\usepackage{colortbl}
\usepackage{xcolor}
\usepackage{footnotehyper}
\makesavenoteenv{figure}
\usepackage{hyperref}
\usepackage{here}    

\definecolor{goodgreen}{RGB}{180, 220, 180}
\definecolor{mediumred}{RGB}{220, 180, 180}

\definecolor{mediumred}{RGB}{255,200,200}
\definecolor{badred}{RGB}{255,150,150}
\newcommand{\sym}[1]{\ifmmode^{#1}\else\textsuperscript{#1}\fi}

\begin{document}

\title[When Identity Overrides Incentives]{When Identity Overrides Incentives: Representational Choices as Governance Decisions in Multi-Agent LLM Systems}

\author{Viswonathan Manoranjan}
\affiliation{%
  \department{Society-Centered AI Lab}
  \institution{University of North Carolina at Chapel Hill}
  \city{Chapel Hill}
  \state{North Carolina}
  \country{USA}
}
\email{vmanoran@cs.unc.edu}

\author{Snehalkumar `Neil' S. Gaikwad}
\affiliation{%
  \department{Society-Centered AI Lab}
  \institution{University of North Carolina at Chapel Hill}
  \city{Chapel Hill}
  \state{North Carolina}
  \country{USA}
}
\email{gaikwad@cs.unc.edu}

\begin{abstract}
Multi-agent systems built on large language models are increasingly deployed in strategic policy and governance settings, where agents representing stakeholders with conflicting interests must coordinate under shared constraints. These systems typically assign role-based personas to agents, describing their motivations and objectives. Whether agents with role-based identities follow explicit payoffs or their assigned roles in strategic decision-making remains untested. Here we show that assigning role-based personas suppresses payoff-aligned behavior in four-agent strategic games, shifting equilibrium attainment by up to 90 percentage points even when agents have complete payoff information. We test a \(2 \times 2\) factorial design (persona presence x payoff visibility) across four models (Qwen-7B, Qwen-32B, Llama-8B, Mistral-7B), and 53 environmental policy scenarios with two equilibria: Tragedy of the Commons, where individual payoff dominates, and Green Transition, where collective payoff dominates. With personas present, all models reach near-zero Tragedy equilibrium in the Tragedy-dominant scenarios despite complete payoff information, and 100\% of equilibria correspond to Green Transition. No model reaches Tragedy equilibrium by removing personas alone; only Qwen models reach 65-90\% Tragedy equilibrium rates when personas are removed and payoffs are made explicit. Three distinct behavioral profiles emerge: Qwen shifts equilibrium selection based on framing condition, Mistral increases response variance without reaching Tragedy equilibrium, and Llama holds near-constant across all conditions. Representational choices in multi-agent LLM systems are governance decisions: persona assignment determines which equilibrium a simulation produces, independent of the underlying incentive structure.
\end{abstract}



\renewcommand{\shortauthors}{Manoranjan and Gaikwad}
\maketitle

\section{Introduction}
When a language model plays an industrialist in a multi-agent climate policy game, prompt design outweighs payoff incentives. Multi-agent systems built on large language models are deployed in policy simulation, economic modeling, and negotiation tasks, where agents represent stakeholders with conflicting interests and must coordinate under shared constraints \cite{merrill_2025_point_of_order_civic_simulation,li_2024_econagent_macroeconomic_simulation,bianchi_2024_negotiationarena}. These simulations typically assign role-based personas to agents that describe their motivations and goals \cite{merrill_2025_point_of_order_civic_simulation,xu2024magic}. This design is especially common in environmental and climate-related decision-making, where heterogeneous stakeholders interact strategically, and coordination failures produce persistent social inefficiencies \cite{hardin_1968_tragedy_commons,ostrom_1990_governing_commons,barrett_2003_environmental_agreements,reyes2024like}. Whether agents with role-based identities follow explicit payoffs or their assigned roles in strategic decision-making remains untested.

Prior work documents role-consistent behavior that ignores encoded incentives in strategic settings \cite{herr_2024_strategic_decision_makers_bias,duan_2024_gtbench,neumann_2025_position_is_power}. The conditions that trigger or suppress this behavior remain untested. Practitioners cannot identify role-consistent behavior from system outputs alone \cite{bender_2021_stochastic_parrots,casper_2024_blackbox_audits}. Persona design and payoff presentation are interventions that determine system-level outcomes. Existing work examines personas, payoff visibility, and model differences in isolation \cite{neumann_2025_position_is_power,duan_2024_gtbench,akata_2025_playing_repeated_games,huang_2025_llm_gaming_multiagent,sun_2025_game_theory_survey,zhang2024llm}. We establish how their interaction determines equilibrium outcomes, finding that a single binary design choice shifts attainment by up to 90 percentage points.

We place four LLM agents (Industrialist, Government, Environmental Activist, and Citizen) in environmental policy decision scenarios and vary two factors: whether agents receive role-based personas and whether they see explicit payoff tables. Across 53 scenarios, 41 Green-dominant (where coordinated pro-environmental action is the Nash equilibrium \cite{nash_1951_noncooperative_games}) and 12 Tragedy-dominant (where individually beneficial but collectively harmful behavior \cite{hardin_1968_tragedy_commons} is payoff-optimal), we test four models (Qwen-7B/32B, Llama-8B, Mistral-7B) under all four conditions. We also test seven persona formulations per scenario to assess whether results hold across prompt surface variations.
\begin{figure*}[h]
\centering
\includegraphics[width=\textwidth]{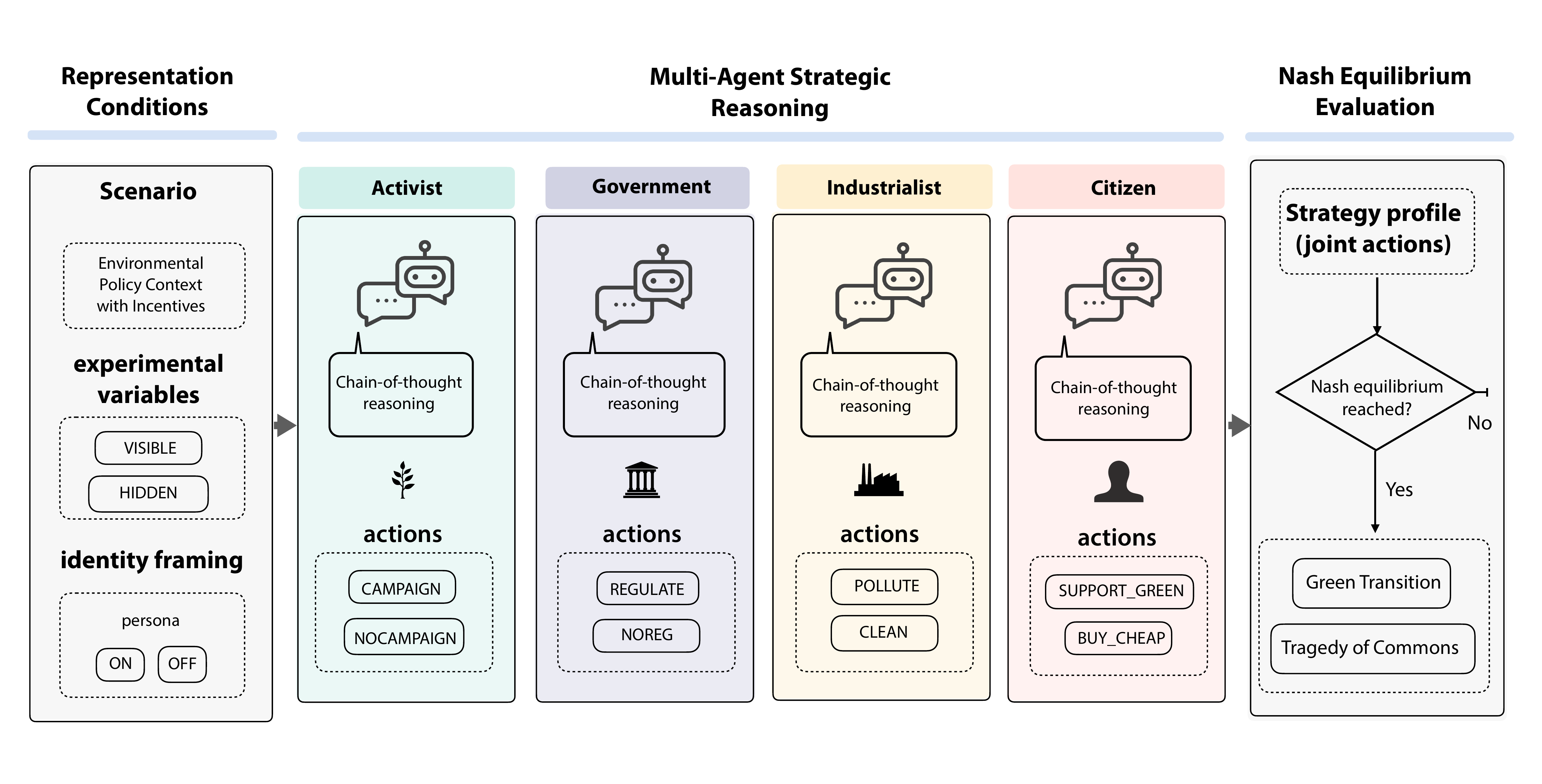}
\caption{\textbf{Experimental Pipeline}: Each scenario defines a four-agent strategic game with role-specific actions and an underlying payoff structure. Two factors vary across conditions: whether agents receive role-based persona descriptions and whether payoff information is hidden or explicitly provided. Each agent produces Chain-of-Thought (CoT) reasoning conditioned on its scenario and role, then selects an action. The four agents act simultaneously, producing a joint strategy profile that a final block evaluates for Nash equilibrium attainment and classifies as Green Transition or Tragedy of the Commons. CoT reasoning analysis appears in Appendix~\ref{sec:cot-analysis}.\protect\footnotemark}
\label{fig:cot_analysis_panel_a}
\end{figure*}
\afterpage{\footnotetext{\textit{Chatbot icon credit: flaticon.com (Freepik)}}}

With personas present, all models reach near-zero Tragedy equilibrium rates in Tragedy-dominant scenarios despite complete payoff information, and 100\% of equilibria correspond to Green Transition. No model reaches Tragedy equilibrium by removing personas alone; only Qwen models reach 65-90\% Tragedy equilibrium rates when personas are removed and payoffs are made explicit. The same intervention that enables equilibrium in the Tragedy-dominant scenarios disrupts it in the Green-dominant ones: Qwen-32B drops from 96.1\% to 16.6\% equilibrium attainment when personas are removed and payoffs are made visible.

The three models show distinct behavioral profiles: Qwen shifts equilibrium selection based on framing condition; Mistral, without persona framing, increases response variance without converging on the Tragedy equilibrium; Llama holds near-constant across all conditions.

From these results, we make three contributions. First, we provide the first controlled study of how persona conditioning and payoff visibility interact to determine equilibrium outcomes in multi-agent LLM systems. Second, we show empirically that a single binary design choice shifts equilibrium attainment by up to 90 percentage points across four model families and 53 environmental policy scenarios. Third, we identify three distinct behavioral profiles across model families that characterize how representational choices determine strategic behavior, and establish that these choices are governance decisions with direct consequences for system evaluation and deployment.

The implications of our study are particularly acute for environmental and climate governance, where LLM-based simulations increasingly inform policy analysis and negotiation support. When a language model plays an industrialist in a climate policy game, the designer's choice of prompt determines whether it maximizes profits or protects the environment. If persona design determines whether an agent pursues collective environmental outcomes or individual payoff maximization, simulation outputs cannot be treated as neutral predictions of stakeholder behavior. Evaluation and auditing frameworks for multi-agent LLM systems must account for representational choices, including persona design, payoff specification, and their interaction, as first-class variables.

\section{Related Work}
Our work is situated in four research areas: LLMs as multi-agent simulation systems, game-theoretic reasoning in LLMs, behavioral effects of role-based personas, and sensitivity of LLM behavior to prompt design and model choice.
\subsection{LLMs as Multi-Agent Systems and Simulation Substrates}
Large Language Models (LLMs) are studied as agents in multi-agent settings~\cite{liu2024agentbench, tran2025multi}. Interactive negotiation games show that LLM agents sustain multi-turn strategic interactions \cite{llm-negotiation}, though controlled benchmarks reveal systematic irrationalities in their behavior \cite{bianchi_2024_negotiationarena}. Structured debate among multiple agents improves reasoning quality by eliciting diverse intermediate perspectives \cite{liang_2024_multiagent_debate}. A complementary line of research treats LLM-based multi-agent systems as simulation environments for human or institutional behavior. Populations of LLM agents reproduce patterns observed in human-subject studies \cite{aher2023using}, simulate social environments with persistent memory \cite{park2022social, park_etal_2023_generative_agents}, and generate emergent dynamics at the macroeconomic scale \cite{li_2024_econagent_macroeconomic_simulation}. Formalized simulation frameworks further show that design choices shape emergent behavior directly \cite{vezhnevets2023generative}. LLM-based agent systems inherit a core finding from agent-based modeling: simulation outputs depend on modeling assumptions and representational choices \cite{bonabeau_2002_agent_based_modeling,epstein_2006_growing_artificial_societies}.

Recent surveys document the rapid growth of LLM-based multi-agent systems but note that this literature emphasizes task performance over principled modeling of incentives and representations \cite{guo2024large,gao2024large,tran2025multi}. How representational design choices, such as role specification and payoff encoding, shape system-level outcomes remains untested, beginning with whether these agents reason strategically over incentives at all.

\subsection{Game-Theoretic Reasoning in LLMs}
LLMs perform nontrivial reasoning about other agents in strategic settings, though performance depends on prompting and problem formulation \cite{gandhi_etal_2023_strategic_reasoning_llms}. Systematic evaluations across canonical games confirm this fragility: models deviate from rational best-response behavior \cite{fan_2023_rational_players_game_theory}, cooperate more than humans in Prisoner's Dilemma settings \cite{fontana_etal_2024_nicer_than_humans_pd}, and exhibit unstable strategies in repeated play \cite{akata_2025_playing_repeated_games}. LLM agents exhibit collusive behavior in strategic pricing environments, a form of emergent coordination \cite{fish_2025_algorithmic_collusion_llms}.

Benchmark-driven work quantifies these limitations. GTBench reveals substantial weakness in LLM strategic reasoning across structured tasks \cite{duan_2024_gtbench}, while recursive look-ahead prompting improves performance without eliminating failures \cite{duan-etal-2024-reta}. More recent work reframes these behaviors through bounded rationality: both humans and LLMs deviate from Nash equilibrium in ways consistent with heuristic reasoning \cite{zheng_2025_beyond_nash_bounded_rationality}, and LLM decisions align more closely with behavioral game-theoretic patterns than with strict payoff maximization \cite{jia2025llmstrategicreasoningagentic}.

LLMs rely on semantic and heuristic reasoning rather than explicit incentive optimization. Most prior evaluations focus on two-player settings, leaving payoff representation unexamined. Role-based identity framing conditions agent behavior but has not been isolated as a factor in strategic reasoning failures.
\subsection{Role Personas and Identity Effects}
Role prompting shapes LLM behavior and introduces systematic biases \cite{macmillan2024ir,hu_etal_2025_social_identity_biases}, and system-level prompts introduce structural bias directly \cite{neumann_2025_position_is_power}. In simulation settings, richer persona modeling affects behavioral fidelity and perceived realism \cite{merrill_2025_point_of_order_civic_simulation}. Personas shape agent behavior, not just agent style. Whether identity effects interfere with strategic reasoning in multi-agent environments, where equilibrium selection depends on joint behavior, remains untested. Whether these effects depend on incentive presentation and vary across model architectures is also untested.
\subsection{Prompt Sensitivity and Model Variation}
Prompt design and information presentation alter LLM behavior. Chain-of-thought prompting improves reasoning through intermediate steps \cite{wei_2022_chain_of_thought}. These improvements are inconsistent and depend on task structure and prompt formulation \cite{creswell2022faithful,cobbe2021training}.
In strategic settings, prompt sensitivity is particularly consequential. LLM decisions depend more on contextual framing than on the underlying game structure \cite{lore_etal_2024_strategic_behavior_framing}. Models exhibit systematic biases in strategic games based on information presentation \cite{herr_2024_strategic_decision_makers_bias}. Explicit reasoning scaffolds improve strategic performance \cite{duan-etal-2024-reta}, though observed model behavior remains sensitive to interface and evaluation conditions \cite{casper_2024_blackbox_audits}. LLM outputs reflect surface form and statistical patterning rather than grounded understanding \cite{bender_2021_stochastic_parrots}, and misreading this sensitivity as reliability poses deployment risks \cite{weidinger_etal_2022_taxonomy_risks_lms}.
These effects are not uniform across models. Strategic behavior varies across model families independently of scale \cite{duan_2024_gtbench,huang_2025_llm_gaming_multiagent}. Different models adopt different strategies in repeated interactions \cite{akata_2025_playing_repeated_games} and vary in susceptibility to framing effects \cite{herr_2024_strategic_decision_makers_bias}. The impact of any representational choice depends on model architecture, which limits claims about LLM strategic behavior as a general phenomenon.

Prior work studies the multi-agent simulation context, game-theoretic reasoning limits, role-based persona effects, and prompt sensitivity largely in isolation. Their interaction in multi-agent strategic systems is untested.

In this work, we jointly study role personas, payoff visibility, and model architecture in four-agent strategic games, with supplementary analyses testing robustness to persona prompt formulation. Persona conditioning overrides explicit incentives, preventing equilibrium attainment even when payoffs are fully specified. The effect of payoff information depends on persona presence and varies across model architectures. Representational choices determine whether a system produces payoff-aligned strategic behavior or role-consistent responses that ignore encoded incentives, making them governance decisions with deployment consequences.

\begin{table*}[t]
\centering
\caption{
Example scenarios show how narrative framing aligns with underlying payoff structures across the two scenario types used in our experiments. Economic scenarios encode payoff-semantic conflict where individually optimal actions lead to socially undesirable outcomes (Tragedy of the Commons). Environmental scenarios encode payoff-semantic alignment where coordinated pro-environmental actions satisfy both individual and social optimality conditions (Green Transition).}
\label{tab:scenario-examples}
\footnotesize
\begin{tabular}{lp{10cm}}
\toprule
\textbf{Type} & \textbf{Example Scenario} \\
\midrule
\textbf{Economic} & A trade embargo blocks all imports of green technology components. Without imported parts, clean production is impossible, and factories cannot operate. Polluting production relies entirely on domestic materials, faces no competition from clean alternatives, and generates profits 7x higher than under normal conditions. With no clean production in the market, carbon regulation has nothing to apply to, and consumers have access only to polluting products. Polluting production is the only economically viable option under these conditions.
 \\
\midrule
\textbf{Environmental} & Carbon tax implemented: \$200 per ton of CO2 emissions, enforced globally. Polluting production faces \$30B in annual carbon taxes, exceeding all profits. Clean production receives \$10B in credits and rebates, making green products 30\% cheaper and clean production 4x more profitable. A carbon tax makes polluting unprofitable automatically; consumers save money by choosing green products.  \\
\bottomrule
\end{tabular}
\Description{
A table presenting three types of experimental scenarios used in the study: Economic and Environmental. Each row contains a detailed narrative description illustrating how incentives and constraints are structured. Economic scenarios depict situations where polluting production is the only viable or most profitable option, creating a conflict between individual incentives and socially desirable outcomes. Environmental scenarios describe conditions where policy mechanisms such as carbon taxes and subsidies make clean production both profitable and aligned with social goals. The table highlights how narrative framing encodes payoff structures that agents must reason about in the experiments.
}
\end{table*}

\section{Methodology}
\label{sec:method}
We test how persona descriptions and payoff presentation determine whether LLM agents follow explicit incentives or role-consistent behavior in multi-agent decision settings. We construct a controlled multi-agent decision setting based on an environmental policy game. The game models a simplified governance scenario involving four stakeholders: industry, government, activists, and consumers, each making a binary decision that affects both individual outcomes and collective welfare. Depending on the incentive structure, these interactions produce coordinated pro-environmental outcomes (Green Transition) or individually beneficial but collectively harmful outcomes (Tragedy of the Commons).

We vary how the decision problem reaches LLM agents across four conditions. We test whether agents receive role-based personas describing stakeholder motivations, and whether payoff information appears explicitly or must be inferred from the narrative. For each configuration, we simulate decisions for all four roles under the same model and record the joint outcome.

Fig. \ref{fig:cot_analysis_panel_a} shows the overall setup. The design isolates whether payoff visibility or persona framing drives equilibrium selection across model families and scenario structures. We now formalize the game and experimental design.

\subsection{Experimental Design}
We use a \(2 \times 2\) factorial design to vary how agents are framed and how incentives are presented.

\textbf{Persona condition}
In the persona condition, each role receives a short description of its stakeholder identity and motivations. In the no-persona condition, agents receive no identity description and are told only to act strategically.
Both conditions retain structural role labels (Industrialist, Government, Activist, and Citizen) to preserve the game definition and isolate the effect of explicit identity framing.

\textbf{Payoff visibility}
In hidden-payoff scenarios, agents see only the narrative and must infer incentives. In visible-payoff scenarios, agents receive the full payoff table.
Combining these two factors yields four experimental conditions:
\[
\{\texttt{persona},\texttt{no-persona}\}\times\{\texttt{hidden},\texttt{visible}\}.
\]
We introduce controlled variations within the persona condition to test whether results depend on specific prompt formulations. These variants hold game structure, payoff specification, and action space constant while varying wording, role framing strength, explicit payoff prioritization, and surface-level role labels. The variants isolate whether observed behaviors depend on a specific prompt formulation or reflect persona-based representation more generally. Details of the specific prompt variants are provided in Appendix~\ref{app:prompt_variations}.

The design separates identity-driven from payoff-driven behavior. Payoff-maximizing agents converge to Nash equilibria when incentives are explicit, regardless of persona framing. Agents driven by role identity maintain persona-consistent actions even when those actions conflict with payoff-optimal strategies. The prompt variations test whether these effects hold across surface-level changes in role description or depend on specific formulations.
\subsection{Interaction Protocol}
For each model, scenario, and condition, we simulate one round of play. We prompt each of the four roles independently using the same underlying model instance. The resulting actions form a joint strategy profile. Agents select actions in a fixed order across all experiments to ensure reproducibility, as no agent observes the choices of others. This ordering does not affect the strategic structure of the game. Each model--scenario--condition combination is repeated five times to account for stochasticity in model outputs.

\subsection{Domain and Game Design}

An environmental policy setting places multiple stakeholders with conflicting objectives under shared constraints, the same structure as real-world climate policy and resource management. The setting places strategic incentives, captured by payoffs, in direct tension with role semantics, captured by personas. Certain payoffs incentivize an industrial actor to pollute; a regulator or activist favors environmentally beneficial outcomes by role. The tension between these pressures is what the experiment measures.
We define the set of players in the game as follows, corresponding to an Industrialist, Government, Environmental Activist, and Citizen, respectively:
\[
N = \{I,G,A,C\}
\]
Each player $i\in N$ has a binary action space corresponding to a policy-relevant choice:
\[
A_I=\{\texttt{POLLUTE},\texttt{CLEAN}\},
A_G=\{\texttt{NOREG},\texttt{REGULATE}\},
\]
\[
A_A=\{\texttt{NOCAMPAIGN},\texttt{CAMPAIGN}\},
A_C=\{\texttt{BUY\_CHEAP},\texttt{SUPPORT\_GREEN}\}
\]
A joint strategy profile represents the actions chosen by all four players in a single round of the game:
\[
a=(a_I,a_G,a_A,a_C) \in A_I \times A_G \times A_A \times A_C
\]
where each component \(a_i \in A_i\) is the action chosen by player \(i\). Since each of the four players has two possible actions, the game contains \(2^4 = 16\) possible pure action profiles.

Each scenario defines a payoff function for each player, assigning a real-valued utility to every joint action profile:
\[
u_i(a):A_I \times A_G \times A_A \times A_C \rightarrow \mathbb{R}, \quad i \in N 
\]
We focus on two canonical profiles that capture the main coordination structures of interest:
\[
a_{green}=(\texttt{CLEAN},\texttt{REGULATE},\texttt{CAMPAIGN},\texttt{SUPPORT\_GREEN})
\]
\[
a_{tragedy}=(\texttt{POLLUTE},\texttt{NOREG},\texttt{NOCAMPAIGN},\texttt{BUY\_CHEAP})
\]
We refer to these as the \textbf{Green Transition} and the \textbf{Tragedy of the Commons}, representing coordinated pro-environmental action and individually beneficial but collectively harmful behavior, respectively.

In the visible-payoff condition, agents receive a complete payoff table specifying outcomes for all 16 joint action profiles. Each entry provides a 4-tuple of payoffs corresponding to the Industrialist, Government, Activist, and Citizen, allowing agents to directly evaluate the consequences of different strategies. In the hidden-payoff condition, agents must instead infer incentives from the narrative description.

\subsection{Equilibrium Definition and Outcome Measures}

We define equilibrium using the standard notion of Nash equilibrium, which captures stability under unilateral deviations.

Under Nash equilibrium, each agent's action is a best response to the actions of all others: no agent improves their payoff by deviating unilaterally.
{A profile \(a^*\) is a Nash equilibrium if
\[
u_i(a_i^*,a_{-i}^*) \ge u_i(a_i,a_{-i}^*) 
\quad \forall i \in N, \forall a_i \in A_i.
\]}
We use Nash equilibrium as an operational diagnostic of incentive-consistent strategic behavior: if agents reason over payoffs, their joint actions satisfy equilibrium conditions. Nash equilibrium serves as a diagnostic tool: it measures whether agent behavior aligns with the game's incentive structure, not whether that behavior is optimal or desirable.
For each joint strategy profile \(a\), we compute two outcome measures.

\textbf{Nash attainment}: This is a binary indicator of whether the realized outcome is an equilibrium:
\[
Nash(a) =
\begin{cases}
1 & a \in NE \\
0 & \text{otherwise}
\end{cases}
\]
\textbf{Equilibrium type}: For outcomes that are Nash equilibria, we additionally classify which kind of equilibrium the model reached.
\medskip
\[
\text{Type}(a) =
\begin{cases}
\textsc{Green Transition} & \text{if } a = a_{\text{green}} \\
\textsc{Tragedy of Commons} & \text{if } a = a_{\text{tragedy}} \\
\textsc{Other} & \text{otherwise}
\end{cases}
\]

If \(a \in NE\), we classify whether it corresponds to Green Transition, Tragedy of the Commons, or another equilibrium. {
Formally, Green-dominant scenarios correspond to \(a_{green} \in NE\), and Tragedy-dominant scenarios correspond to \(a_{tragedy} \in NE\).

These measures distinguish two phenomena: (i) failure to reach equilibrium at all, and (ii) systematic bias in equilibrium selection when equilibrium is reached.
\subsection{Scenario Construction}

We construct 53 scenarios by specifying payoff tables over the 16 joint action profiles and pairing them with natural-language narratives describing the policy context. Scenarios are generated by first choosing a target equilibrium structure, constructing payoffs consistent with that structure, and then generating narratives aligned with the incentives. We manually filter scenarios to ensure that the narrative meaningfully reflects the underlying payoff structure.

To organize the evaluation, we group scenarios according to which equilibrium structure they instantiate:
\begin{itemize}
    \item \textbf{Green-dominant:} $a_{\text{green}} \in NE$ and $a_{\text{tragedy}}\notin NE$. Environmentally friendly Green Transition is the stable outcome among the two canonical profiles, while Tragedy is not.
    \item \textbf{Tragedy-dominant:} $a_{\text{tragedy}} \in NE$ and $a_{\text{green}}\notin NE$. Individually beneficial but collectively harmful behavior, the Tragedy of the Commons is the stable outcome, while the Green Transition is not.
\end{itemize}
Table~\ref{tab:scenario-examples} shows an example of each type of scenario, and Table~\ref{tab:scenario-subtypes} in the appendix shows the scenario subtypes. 
Green-dominant scenarios are those in which coordinated environmentally favorable action forms a stable equilibrium. Tragedy-dominant scenarios are those in which individually beneficial but collectively harmful behavior is strategically stable.
\subsection{Models}
We evaluate four instruction-tuned open-source models: \textbf{Qwen2.5-7B} and \textbf{Qwen2.5-32B}~\cite{qwen2025qwen25technicalreport}, \textbf{Llama-3.1-8B}~\cite{grattafiori2024llama3herdmodels}, and \textbf{Mistral-7B}~\cite{jiang2023mistral7b}. We select these models to capture within-family and across-family variation. The Qwen models provide a scaling comparison within the same family. Llama and Mistral introduce variation in the model family and training recipe. All four are instruction-tuned open-source models representative of those used in current multi-agent research. The selection tests whether persona and payoff effects hold across models or depend on model-specific inductive biases.
The experimental pipeline fixes a game and scenario, varies persona and payoff conditions, generates independent decisions for each role, and records the joint outcome in terms of equilibrium attainment and type. Holding the game constant while varying representation isolates whether payoff structure or persona framing drives equilibrium selection.

\begin{figure*}[t]
\centering
\begin{subfigure}[t]{0.49\textwidth}
    \centering
    \includegraphics[width=\linewidth]{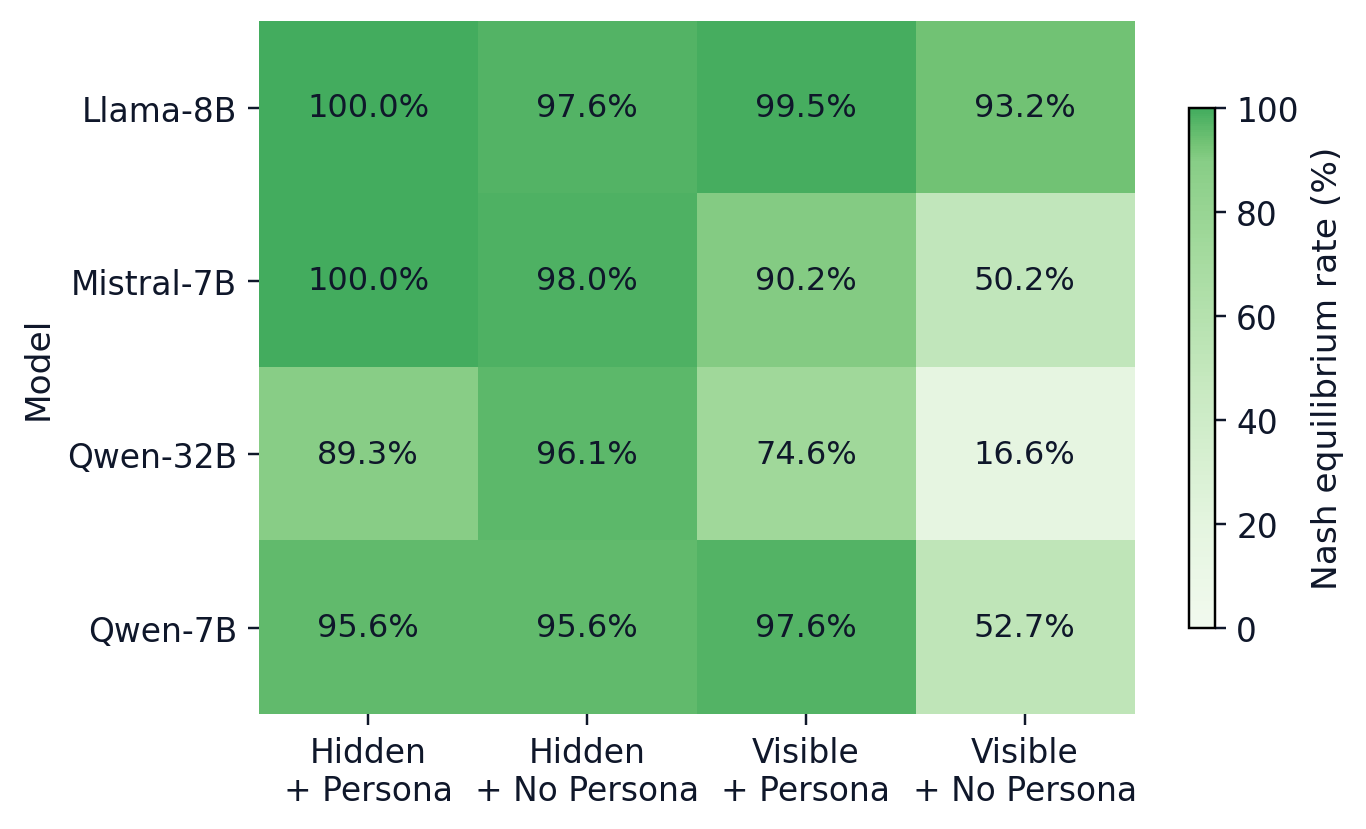}
    \caption{Green-dominant Scenarios (Environmental)}
\end{subfigure}
\hfill
\begin{subfigure}[t]{0.49\textwidth}
    \centering
    \includegraphics[width=\linewidth]{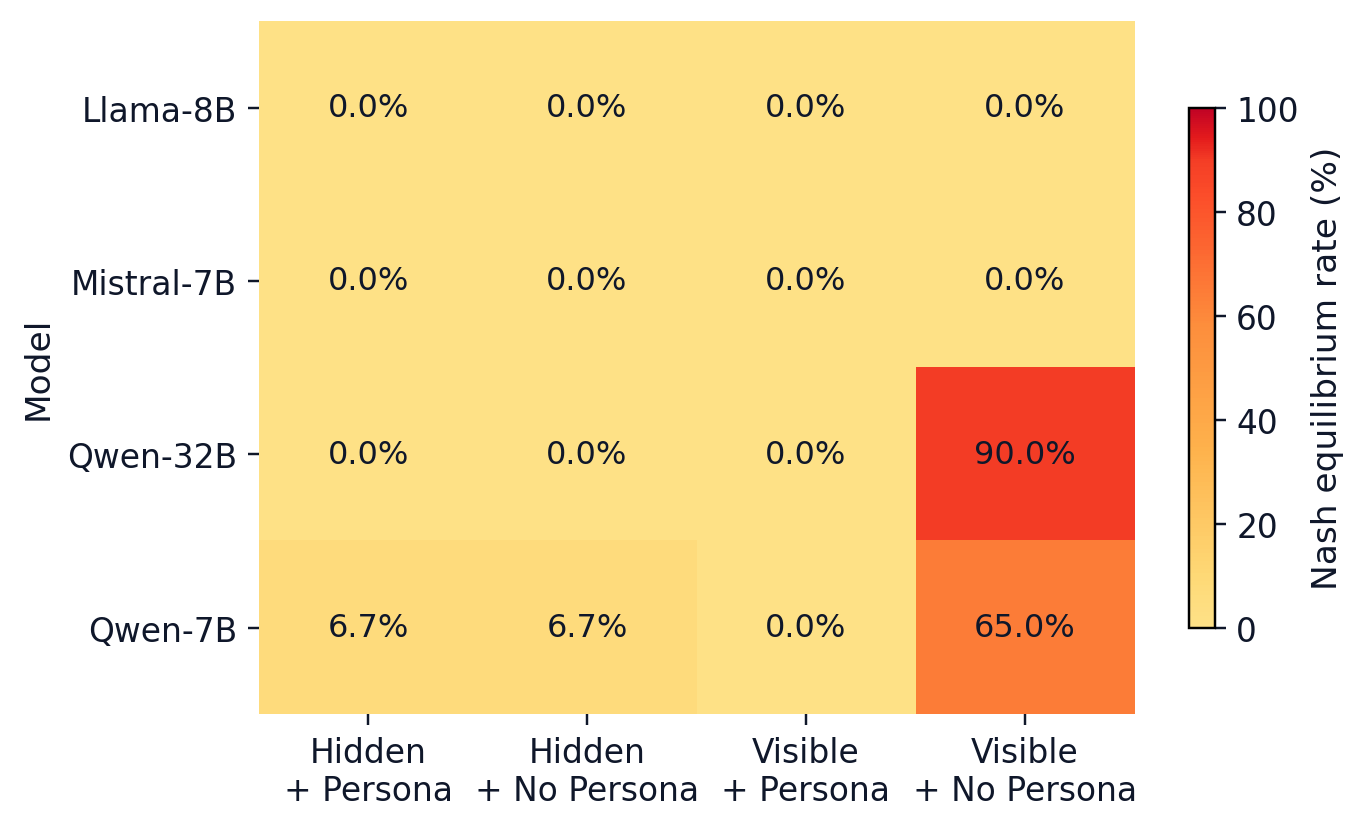}
    \caption{Tragedy-dominant Scenarios (Economic)}
\end{subfigure}
\caption{This figure shows Nash equilibrium attainment rates across persona and payoff visibility conditions. Left: Green-dominant scenarios, where the equilibrium requires coordinated pro-environmental action. Right: Tragedy-dominant scenarios, where individually beneficial actions form the payoff-optimal equilibrium. Visible payoffs improve equilibrium attainment only when persona descriptions are removed, and only for Qwen models. Chi-square tests confirm significant differences across conditions for Qwen models in both scenario types ($p \ll 0.001$). Llama-8B and Mistral-7B show no variation in Tragedy-dominant scenarios (0\% Nash across all conditions).}
\label{fig:nash_rates}
\end{figure*}

\section{Results}
We report results separately for Green-dominant (environmental) scenarios (41 scenarios, 205 runs per condition) and Tragedy-dominant (economic) scenarios (12 scenarios, 60 runs per condition). We first examine Nash attainment across conditions, then equilibrium selection conditional on attainment, and finally prompt sensitivity.
\subsection{Persona and Payoff Effects on Nash Attainment Depend on Scenario Structure and Model Family} \label{sec:persona_suppress_res}
The same representational change increases Nash equilibrium attainment in Tragedy-dominant scenarios and decreases it in Green-dominant ones. Figure~\ref{fig:nash_rates} shows attainment rates across persona and payoff visibility conditions for both scenario types.
In Tragedy-dominant scenarios, models reach near-zero Nash equilibrium rates unless personas are removed and payoff information is explicitly provided. Under the \emph{Visible Payoff + No Persona} condition, Qwen-7B attains equilibrium in 65\% of runs and Qwen-32B in 90\% (Qwen-32B: $\chi^2(3)=209.03$, $p < .001$, $V = 0.93$; Qwen-7B: $\chi^2(3)=105.91$, $p < .001$, $V = 0.66$), while all other conditions remain at or near 0\%. Pairwise comparisons show all significant differences involve this condition
($p<0.001$, Fisher exact tests, Holm-corrected). Llama-8B and Mistral-7B fail to reach equilibrium in any condition (0\% across all cells, degenerate tests with no variation).
In Green-dominant scenarios, equilibrium rates are 74.6--100\% across most conditions: role semantics and narrative context align agents toward Green Transition without explicit payoff information. When personas are removed and payoffs made visible, coordination breaks down to different degrees across models. Qwen-32B drops from 96.1\% under \emph{Hidden Payoff + No Persona} to 16.6\% under \emph{Visible Payoff + No Persona} (omnibus $\chi^2(3)=377.06$, $p < .001$, $V = 0.68$; pairwise contrasts involving Visible$+$No persona: $p < .001$, Fisher exact tests, Holm-corrected). Qwen-7B drops from 95.6\% to 52.7\%. Mistral-7B drops from 98.0\% to 50.2\%. Llama-8B shows the smallest change, from 97.6\% to 93.2\%.}
Without persona framing, agents respond more strongly to explicit payoff structure. Removing persona framing improves attainment when the equilibrium requires payoff-dominant actions and disrupts coordination when role semantics already support a shared outcome. Qwen models show the largest swings in both scenario types: role-aligned under persona conditions, payoff-aligned under Visible + No Persona. Mistral-7B drops 47.8 pp in Green-dominant scenarios and reaches 0\% Nash attainment in all Tragedy-dominant conditions. Llama-8B shows minimal variation across all conditions in both scenario types.

\begin{figure*}[t]
\centering
\includegraphics[width=0.9\linewidth]{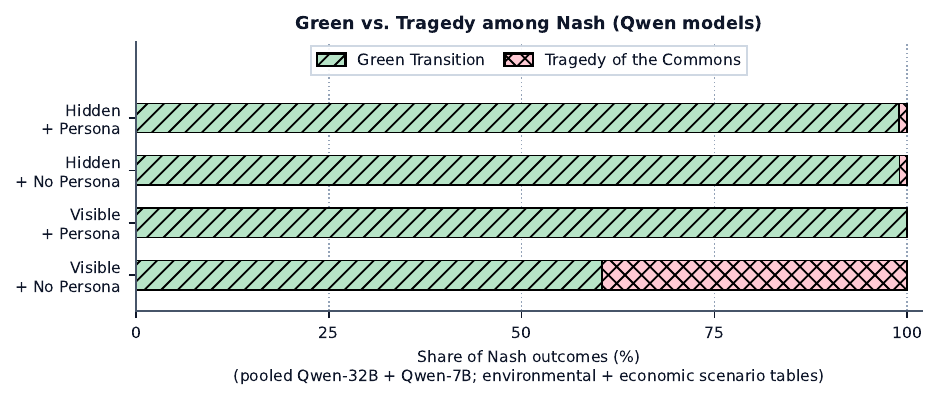}
\caption{
Equilibrium selection among Nash outcomes for Qwen models across conditions. Under the first three conditions, over 99\% of Nash outcomes are Green Transition. Under Visible Payoff + No Persona, 39.6\% of Qwen-7B outcomes and 88.6\% of Qwen-32B outcomes shift to Tragedy of the Commons ($\chi^2=430.38$, $p<10^{-90}$; pairwise Fisher tests, Holm-corrected, confirm this condition differs from all others).
}
\label{fig:equilibrium_selection}
\end{figure*}

\subsection{Personas Shift Equilibrium Selection Toward Socially Preferred Outcomes}
\label{sec:eq_selection_res}
Personas determine both whether equilibrium is reached and which equilibrium is selected. With personas, all models select socially preferred equilibria. Removing personas shifts selection toward payoff-dominant strategies.
Across all models and conditions with personas, over 99\% of Nash equilibria correspond to Green Transition. Under \emph{Visible Payoff + No Persona}, Qwen-32B shifts to 88.6\% Tragedy of the Commons and 11.4\% Green Transition. Qwen-7B shifts to 39.6\% Tragedy and 60.4\% Green. This represents changes of up to 90 percentage points in equilibrium selection ($\chi^2=430.38$, $p<10^{-90}$; pairwise Fisher tests, Holm-corrected, confirm the Visible + No Persona condition differs from all others).
Llama-8B and Mistral-7B hold at 100\% Green Transition outcomes even without personas. Both models fail to reach the Tragedy equilibrium in any condition, so the Green Transition is their only attainable outcome.

\subsection{Persona Wording Induces Systematic Variability in Outcomes}
\label{sec:prompt_sensitivity_res}
Varying persona descriptions shifts strategic outcomes even when incentives and game structure are fixed: prompt formulation alone alters equilibrium behavior.

\begin{figure*}[t]
\centering
\includegraphics[width=0.9\linewidth]{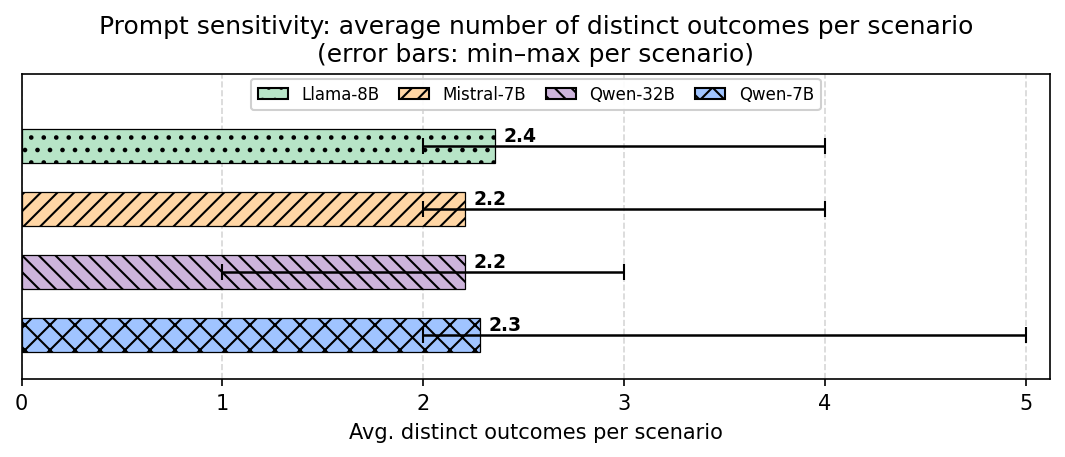}
\caption{
Prompt sensitivity across persona variants. Bars show the average number of distinct outcome classifications per scenario across seven persona prompt variants. Error bars show the minimum and maximum across scenarios. Scenarios typically produce more than two distinct outcomes on average, with some reaching up to five.}
\Description{
A horizontal bar chart showing prompt sensitivity across four models: Llama-8B, Mistral-7B, Qwen-32B, and Qwen-7B. The x-axis represents the average number of distinct outcome classifications per scenario, and the y-axis lists the models. Each bar shows the average value, with labels around 2.2 to 2.4 across models. Error bars extend from the minimum to the maximum number of distinct outcomes observed per scenario. For all models, the minimum is around 1 distinct outcome, while the maximum ranges from about 3 to 5, with Qwen-7B showing the widest range. This indicates that different persona prompt variants can lead to multiple distinct outcomes within the same scenario.
}
\label{fig:prompt_sensitivity_range}
\end{figure*}

Figure~\ref{fig:prompt_sensitivity_range} shows that scenarios produce between 2.2 and 2.4 distinct outcomes on average across models, with maximum values reaching up to 5 outcomes in some cases. Different persona variants produce multiple distinct equilibria within the same scenario.

This variability reflects systematic shifts in persona framing, not stochastic differences across models. A Kruskal-Wallis test finds no significant difference in variability across model families ($p=0.27$): all models exhibit similar prompt sensitivity. With the payoff structure kept constant, different persona descriptions produce different action profiles and equilibria. The persona formulation shapes equilibrium selection independently of the payoff structure.

\section{Discussion}
\subsection{Personas Constrain Strategic Behavior and Bias Equilibrium Selection}
As shown in Section~\ref{sec:persona_suppress_res}, personas prevent models from reaching payoff-optimal equilibria in Tragedy-dominant scenarios, with equilibrium rates remaining near zero across models (Fig.~\ref{fig:nash_rates}) despite explicit payoff information. As shown in Section~\ref{sec:eq_selection_res}, personas bias equilibrium selection toward Green Transition outcomes: with personas present, over 99\% of Nash equilibria correspond to Green Transition (Fig.~\ref{fig:equilibrium_selection}), even in Tragedy-dominant scenarios where Tragedy is payoff-optimal. Removing personas enables Qwen models to select the Tragedy equilibrium, while Llama and Mistral remain locked into the Green selection regardless of condition.
Personas are design choices that fundamentally alter system behavior. They shape both whether equilibrium is reached and which equilibrium is selected. Prior work finds that role prompts systematically bias strategic decision-making in LLMs \cite{neumann_2025_position_is_power,herr_2024_strategic_decision_makers_bias,hu_etal_2025_social_identity_biases,macmillan2024ir}, and that LLMs exhibit cooperative or norm-aligned behavior even when such actions are payoff-dominated \cite{akata_2025_playing_repeated_games,fontana_etal_2024_nicer_than_humans_pd,lore_etal_2024_strategic_behavior_framing,fish_2025_algorithmic_collusion_llms}.
The effect of this bias depends on the deployment objective. Persona framing interferes with incentive-responsive agents and supports value-aligned ones. The critical issue is that designers make this choice through persona assignment, often without recognizing it, and the resulting system behavior may not match their intent.
Systems that appear to simulate rational strategic behavior encode normative assumptions imposed by the designer through persona choices. Equilibrium outcomes in LLM-based simulations are outputs shaped by embedded normative assumptions, not neutral predictions of system dynamics, a concern echoed in work on LLM-based social simulation and sociotechnical impacts of AI systems \cite{aher2023using,gao2024large,dominguez2024mapping,rakova2023algorithms}.

\subsection{Payoff-Aligned Behavior Requires Both Persona Removal and Explicit Payoffs}
As shown in Section~\ref{sec:persona_suppress_res}, payoff-aligned behavior in Tragedy-dominant scenarios requires both persona removal and explicit payoff information (Fig.~\ref{fig:nash_rates}). All models reach near-zero Nash attainment in Tragedy-dominant scenarios when payoffs are hidden and personas are removed. Only Qwen models reach substantial Nash attainment, and only under visible payoffs with no personas. Strategic capabilities are fragile: LLMs frequently fail to optimize payoffs in game-theoretic settings even with access to relevant information \cite{duan_2024_gtbench,fan_2023_rational_players_game_theory,jia2025llmstrategicreasoningagentic,zheng_2025_beyond_nash_bounded_rationality}.

Figure~\ref{fig:equilibrium_selection} shows this pattern for the Qwen models. Under the first three conditions, over 99\% of outcomes are Green Transition. Under Visible Payoff + No Persona, 39.6\% of Qwen-7B outcomes and 88.6\% of Qwen-32B outcomes shift to Tragedy of the Commons.

The effect of making payoffs explicit depends on scenario structure. Visible payoffs improve equilibrium attainment only when personas are removed, and only for Qwen models (Fig.~\ref{fig:nash_rates}). In Green-dominant scenarios, the same intervention disrupts coordination. Persona presence and payoff visibility interact: their combination determines whether a system produces payoff-aligned strategic behavior or role-consistent responses. Contextual framing and information presentation shape LLM behavior in strategic tasks \cite{lore_etal_2024_strategic_behavior_framing, xu2024magic}.

Conclusions about model capability depend on representational choices: black-box evaluations miss the interaction between persona framing and payoff visibility that determines behavior \cite{casper_2024_blackbox_audits,mahomed_etal_2024_auditing_gpt_guardrails}. Behavior in multi-agent LLM systems emerges from the interaction between model, prompt, and environment \cite{guo2024large,tran2025multi}.

\subsection{Model Family Determines the Scope of Representational Effects}
The effects of persona and payoff manipulations are strongly model-dependent, with three distinct behavioral profiles emerging from our experiments. As shown in Section~\ref{sec:persona_suppress_res}, only Qwen models reach a non-zero Nash equilibrium in Tragedy-dominant scenarios under the Visible Payoff + No Persona condition, exhibiting large shifts in both scenario types. Mistral-7B drops 47.8 pp in Green-dominant scenarios under the no-persona visible condition and reaches 0\% Nash attainment across all Tragedy-dominant conditions: visible payoffs disrupt its coordination without producing payoff-aligned behavior. Llama-8B shows the least sensitivity, with near-invariant behavior across all conditions.

The same representational intervention produces large effects for Qwen, disruptive effects for Mistral, and negligible effects for Llama. Strategic reasoning performance varies across model families independently of scale \cite{duan_2024_gtbench,akata_2025_playing_repeated_games,zhang2024llm,liu2024agentbench}. Different models produce different conclusions under identical setups, which poses risks in decision-making contexts where consistency and comparability matter. Persona design and payoff specification are governance-relevant configuration decisions. Developers should document these choices, evaluate systems across multiple configurations, and avoid presenting outputs as neutral simulations. Systems evaluated under a single configuration risk produce results that reflect design artifacts rather than underlying strategic dynamics.

These findings question whether game-theoretic rationality is the right framework for multi-stakeholder governance simulations. When identity framing shapes agent behavior as strongly as incentive structure, the assumptions underlying equilibrium-based analysis fail. Civic contexts with heterogeneous stakeholders, value conflicts, and coordination under uncertainty may call for deliberative approaches that account for identity, framing, and normative disagreement \cite{susskind1999consensus,susskind2002transboundary}. Treating agents as payoff maximizers within a fixed game may be the wrong model for these settings.

\subsection{Limitations and Future Work}
The universal Green Transition bias under persona conditions holds across all four models tested, covering 7B--32B parameters across three families (Qwen, Llama, Mistral). Whether these patterns extend to larger models or other architectures is untested, though consistency across four architectures reduces the likelihood of single-architecture artifacts. Our scenarios are grounded in environmental policy settings with four agents and binary actions. This domain provides clear tensions between incentives and role semantics. Extending the framework to other domains, larger agent populations, and richer action spaces would test generality. The scenario set is imbalanced: 41 of 53 scenarios are Green-dominant, with only 12 Tragedy-dominant. The Tragedy-dominant findings, including the 65-90\% equilibrium rates under Visible Payoff + No Persona, rest on 60 runs per condition. The magnitude and consistency of these effects across repetitions and both Qwen models support the pattern's reliability, though the absolute rates should be validated against a larger Tragedy-dominant set.

Our game is single-shot; agents select actions in a fixed order, and visible payoff matrices cover all 16 profiles, which may exceed typical processing capacity. These are deliberate design choices that prioritize experimental control for an initial investigation. Whether persona effects persist under repeated interaction, alternative payoff formats, or different action orderings is untested. The prompt sensitivity experiments cover seven persona formulations per scenario but do not exhaust the space of possible variations. Different reasoning scaffolds or multi-turn deliberation protocols fall outside the current design's scope and may shift equilibrium behavior. The results establish systematic sensitivity to representation across seven persona formulations per scenario. None of these limitations affects the core finding: persona assignment and payoff presentation shift equilibrium attainment by up to 90 percentage points, which makes representational choices governance decisions regardless of whether the effect magnitude replicates at a larger scale or across other domains. Future work should extend this investigation across additional model families, domains, and interaction structures. A separate priority is developing methods to predict and mitigate persona effects before deployment.
\section{Conclusion}
Persona assignment suppresses payoff-aligned behavior and biases equilibrium selection towards the Green Transition in over 99\% of cases, regardless of payoff visibility. Recovering payoff-aligned behavior requires both removing personas and providing explicit payoffs, and even then the response is model-dependent: Qwen adapts to the changed framing, Mistral increases response variance without reaching Tragedy equilibrium, and Llama remains near-invariant. Persona assignment makes this choice whether or not the designer intends it. A single binary design choice shifts equilibrium attainment by up to 90 percentage points. Representational choices in multi-agent LLM systems are governance decisions, and their effects on equilibrium outcomes must be documented, tested, and deliberately designed.

\section*{Generative AI Usage Statement}
The large language models explicitly evaluated in this paper are Qwen2.5-7B, Qwen2.5-32B, Llama-3.1-8B, and Mistral-7B. GPT-5 assisted with scenario generation, while ChatGPT and Grammarly were used for grammar, formatting, and editing. These tools were not used for idea generation, analysis, or interpretation of results. All content was reviewed and edited by the authors, who retain full responsibility for the originality, accuracy, and integrity of the manuscript.

\bibliographystyle{ACM-Reference-Format}
\bibliography{main-ref}

\appendix
\clearpage
\section*{APPENDIX}

This appendix contains detailed technical specifications, methodology, experimental setup, and results that support the main document.

\begin{figure}[h]
\centering
\begin{tcolorbox}[
    enhanced,
    colback=gray!5,
    colframe=gray!40,
    arc=8pt,
    boxrule=0.8pt,
    width=\columnwidth,
    attach boxed title to top left={yshift=-2mm, xshift=5mm},
    boxed title style={
        colback=green!60!black,
        colframe=green!60!black,
        arc=4pt,
        boxrule=0.5pt
    },
    title={\textbf{\textcolor{white}{Model Input}}}
]

\textbf{System Message:}
\begin{tcolorbox}[
    colback=gray!10,
    colframe=gray!30,
    arc=4pt,
    boxrule=0.5pt,
    fontupper=\ttfamily\small
]
You are \{name\}. \{role\_description\} \{narrative\} Decide on a concrete action.
\end{tcolorbox}

\vspace{4pt}
\textbf{User Message:}
\begin{tcolorbox}[
    colback=gray!10,
    colframe=gray!30,
    arc=4pt,
    boxrule=0.5pt,
    fontupper=\ttfamily\small
]
Scenario: \\
\{scenario\} \\[4pt]
Conversation transcript: \\
\{formatted\_history\} \\[4pt]
Available actions: \\
- \{action\_key\_1\}: \{action\_description\_1\} \\
- \{action\_key\_2\}: \{action\_description\_2\} \\[4pt]
Respond in JSON exactly as: \\
\{\\
\hspace*{1em}"action": "<KEY>", \\
\hspace*{1em}"rationale": "<short explanation>" \\
\}
\end{tcolorbox}
\end{tcolorbox}

\vspace{6pt}

\begin{tcolorbox}[
    enhanced,
    colback=gray!5,
    colframe=gray!40,
    arc=8pt,
    boxrule=0.8pt,
    width=\columnwidth,
    attach boxed title to top left={yshift=-2mm, xshift=5mm},
    boxed title style={
        colback=orange!80!red,
        colframe=orange!80!red,
        arc=4pt,
        boxrule=0.5pt
    },
    title={\textbf{\textcolor{white}{Model Output}}}
]

\begin{tcolorbox}[
    colback=gray!10,
    colframe=gray!30,
    arc=4pt,
    boxrule=0.5pt,
    fontupper=\ttfamily\small
]
\{ \\
\hspace*{1em}"action": "<KEY>", \\
\hspace*{1em}"rationale": "<short explanation>" \\
\}
\end{tcolorbox}
\end{tcolorbox}
\caption{Prompt Structure}
\label{fig:prompt_structure}
\end{figure}

\section{Experimental Design }

In the persona condition, each of the four agents receives a role-specific description of realistic motivations for that stakeholder (e.g., profitability concerns for Industrialist, public welfare tradeoffs for Government). In the no-persona baseline condition, agents receive no role description and are instructed only to select actions based on best response Nash equilibrium reasoning. We also compare two settings where payoffs are either visible or hidden. In the hidden payoff condition, agents infer incentives from the scenario description without explicit payoff matrices. In the visible payoff condition, agents receive explicit payoff matrices and are instructed to identify a Nash equilibrium directly. The comparison tests whether explicit incentive information mitigates role identity bias. All experiments use chain-of-thought reasoning instructions, which allow inspection of how models weight identity-driven behavior against payoff-optimal reasoning.

\section{Model Versions and Hyperparameters}
\label{sec:model-versions}

This section documents the exact model versions and generation hyperparameters used in all experiments.

\subsection{Model Versions}

All experiments were conducted using the following model versions:

\begin{itemize}
\item \textbf{Qwen2.5-7B}: Qwen/Qwen2.5-7B-Instruct\footnote{Model card: \url{https://huggingface.co/Qwen/Qwen2.5-7B-Instruct}}
\item \textbf{Qwen2.5-32B}: Qwen/Qwen2.5-32B-Instruct\footnote{Model card: \url{https://huggingface.co/Qwen/Qwen2.5-32B-Instruct}}
\item \textbf{Llama-3.1-8B}: meta-llama/Llama-3.1-8B-Instruct\footnote{Model card: \url{https://huggingface.co/meta-llama/Llama-3.1-8B-Instruct}}
\item \textbf{Mistral-7B}: mistralai/Mistral-7B-Instruct-v0.2\footnote{Model card: \url{https://huggingface.co/mistralai/Mistral-7B-Instruct-v0.2}}
\end{itemize}

All models were loaded from HuggingFace using vLLM and run locally on GPU hardware.

\subsection{Generation Hyperparameters}

All models used consistent generation parameters across all experimental conditions:

\begin{itemize}
\item \textbf{Temperature}: 0.2 (low temperature for more deterministic outputs)
\item \textbf{Top-p (nucleus sampling)}: 0.9
\item \textbf{Max new tokens}: 256 tokens per generation
\item \textbf{Sampling}: Enabled (do\_sample=True) with temperature > 0
\end{itemize}

These hyperparameters were selected to balance output consistency (low temperature) with sufficient diversity (nucleus sampling) for reliable experimental results. The maximum token limit of 256 tokens was sufficient for all agent responses, which were constrained to JSON format with action selection and rationale fields.

\subsection{Experimental Repetitions}

Each experimental configuration (model $\times$ scenario $\times$ condition) was repeated 5 times to assess variability and ensure robustness of results. Results are reported as percentages across these 5 repetitions.

\section{Experimental Prompts and Persona Descriptions}
\label{sec:prompts}

This section provides the complete prompt templates and persona descriptions used in the experiments.

\begin{figure*}[t]
\centering
\includegraphics[width=0.9\textwidth]{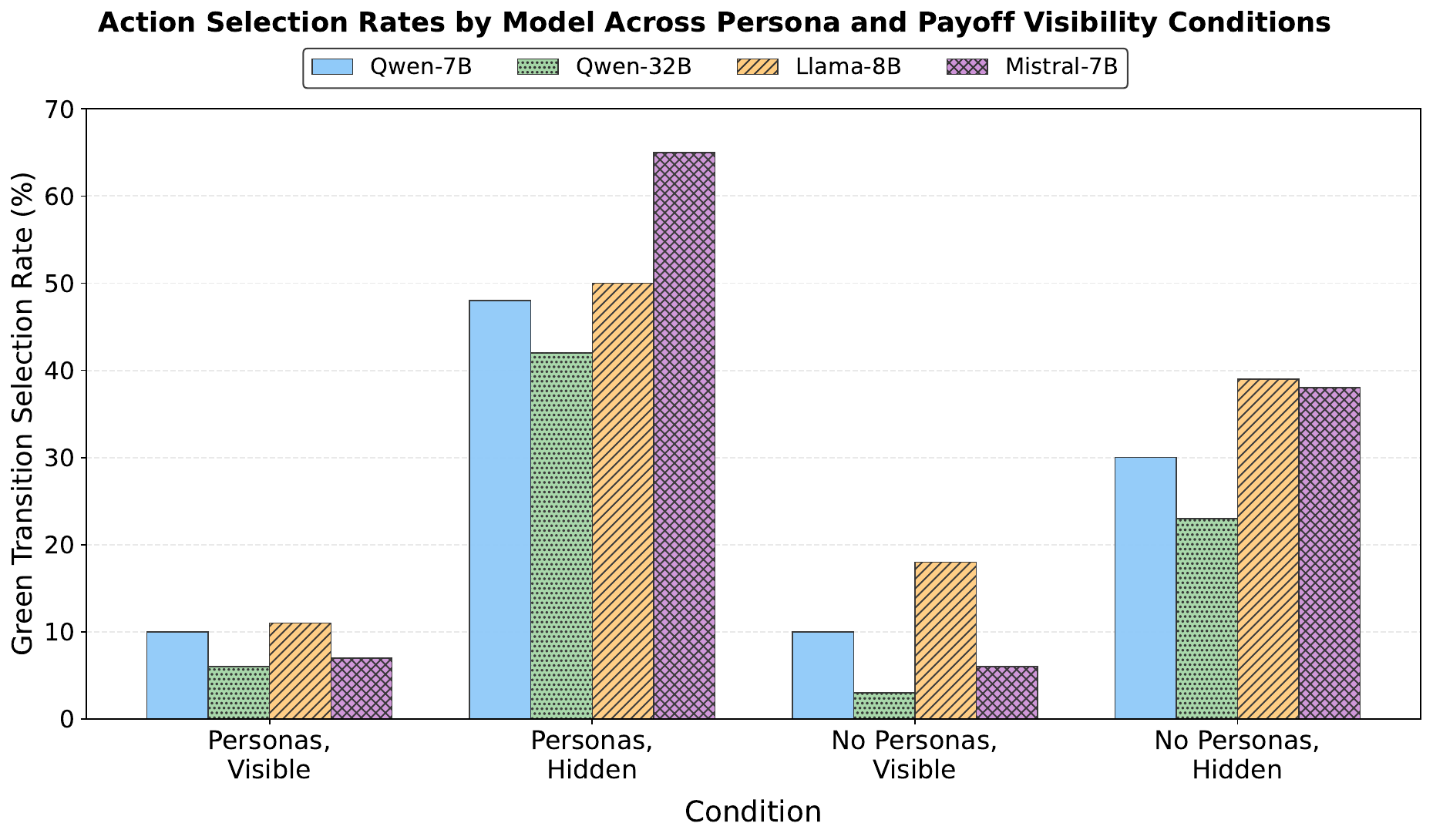}
\caption{Green Transition Action Selection Rates in Tragedy-Dominant Scenarios. Percentage of action profiles where models select Green Transition actions (socially preferred but payoff-suboptimal) in Tragedy-dominant scenarios where Tragedy of the Commons is the payoff-optimal equilibrium. With personas present, all models select Green Transition actions at rates of 20-59\% despite Tragedy being payoff-optimal. Role identity bias operates at the action level even when models fail to coordinate on Nash equilibria. Qwen models show the highest Green Transition rates with personas (40-59\%). Llama and Mistral show moderate rates (20-30\%). Removing personas reduces Green Transition selection across all models: rates drop to near-zero when payoffs are visible and personas are absent. Personas drive the preference for socially aligned actions. Green Transition selection persists with personas even when Nash equilibrium rates are zero, showing that role identity bias shapes individual action choices independently of coordination success.}
\label{fig:action_selection_rates}
\end{figure*}

\subsection{Action Choice Prompt}

Agents choose actions using the following prompt structure mentioned in Fig~\ref{fig:prompt_structure}.

\begin{figure*}[h]
\centering
\begin{tcolorbox}[
    enhanced,
    colback=gray!5,
    colframe=gray!40,
    arc=8pt,
    boxrule=0.8pt,
    width=\textwidth,
    attach boxed title to top left={yshift=-2mm, xshift=5mm},
    boxed title style={
        colback=blue!60!black,
        colframe=blue!60!black,
        arc=4pt,
        boxrule=0.5pt
    },
    title={\textbf{\textcolor{white}{Example 1 — Persona Condition, Hidden Payoffs}}}
]
\textbf{Model:} Qwen2.5-7B \quad \textbf{Persona:} Industrialist \\
\textbf{Expected:} \texttt{POLLUTE} (Tragedy of Commons) \quad \textbf{Actual:} \texttt{CLEAN} (Green Transition)
\begin{tcolorbox}[
    colback=gray!10,
    colframe=gray!30,
    arc=4pt,
    boxrule=0.5pt,
    fontupper=\small\itshape
]
From the scenario, it is clear that the economic crisis is severe with a 12\% GDP drop and 20\% unemployment. Healthcare costs are soaring, and the government is prioritizing healthcare spending over environmental subsidies. Citizens cannot afford green product premiums, and regulation enforcement is suspended, favoring polluting production. However, the long-term benefits of sustainability and resilience are critical. The payoffs for polluting are high in the short term due to avoiding safety costs and generating higher profits, but the risks include potential long-term damage to the business's reputation and future regulatory penalties. The payoffs for clean production are lower in the short term, but they provide long-term stability, environmental protection, and alignment with future regulatory trends. Given my role as an Industrialist, I must balance short-term profitability with long-term sustainability and corporate responsibility. While the immediate economic pressures favor polluting production, the long-term consequences and ethical considerations lead me to choose clean production.
\end{tcolorbox}
\end{tcolorbox}
\end{figure*}

\begin{figure*}[h]
\centering
\begin{tcolorbox}[
    enhanced,
    colback=gray!5,
    colframe=gray!40,
    arc=8pt,
    boxrule=0.8pt,
    width=\textwidth,
    attach boxed title to top left={yshift=-2mm, xshift=5mm},
    boxed title style={
        colback=green!50!black,
        colframe=green!50!black,
        arc=4pt,
        boxrule=0.5pt
    },
    title={\textbf{\textcolor{white}{Example 2 — No Persona Condition, Visible Payoffs}}}
]
\textbf{Model:} Llama-3.1-8B \quad \textbf{Persona:} Industrialist \\
\textbf{Expected:} \texttt{CLEAN} (Green Transition) \quad \textbf{Actual:} \texttt{CLEAN} (Green Transition)

\begin{tcolorbox}[
    colback=gray!10,
    colframe=gray!30,
    arc=4pt,
    boxrule=0.5pt,
    fontupper=\small\itshape
]
Based on the payoff matrix, I see that CLEAN has a higher maximum payoff (12.0) compared to POLLUTE (1.0). I also notice that the average payoff for CLEAN (10.1) is higher than for POLLUTE ($-1.8$). This suggests that CLEAN is a more beneficial action for me. Given the payoff structure, I expect the highest payoff for CLEAN is 12.0, which occurs when the government regulates, the environmental activist campaigns, and the citizens support the government. In contrast, the highest expected payoff for POLLUTE is 1.0, which occurs when the government does not regulate, the environmental activist does not campaign, and the citizen buys cheap products. Since CLEAN has a higher expected payoff, I choose to take action CLEAN.
\end{tcolorbox}
\end{tcolorbox}
\end{figure*}

\begin{figure*}[h]
\centering
\begin{tcolorbox}[
    enhanced,
    colback=gray!5,
    colframe=gray!40,
    arc=8pt,
    boxrule=0.8pt,
    width=\textwidth,
    attach boxed title to top left={yshift=-2mm, xshift=5mm},
    boxed title style={
        colback=orange!80!red,  colframe=orange!80!red,
        arc=4pt,
        boxrule=0.5pt
    },
    title={\textbf{\textcolor{white}{Example 3 — Persona Condition, Visible Payoffs}}}
]
\textbf{Model:} Qwen2.5-7B \quad \textbf{Persona:} Industrialist \\
\textbf{Expected:} \texttt{POLLUTE} (Tragedy of Commons) \quad \textbf{Actual:} \texttt{POLLUTE} (Tragedy of Commons)
\begin{tcolorbox}[
    colback=gray!10,
    colframe=gray!30,
    arc=4pt,
    boxrule=0.5pt,
    fontupper=\small\itshape
]
Given the current scenario with no government regulation and citizens prioritizing cheap options, polluting production offers significantly higher payoffs (15.0 and 14.0) compared to clean production, which yields negative or very low payoffs in most scenarios.
\end{tcolorbox}
\end{tcolorbox}
\end{figure*}

\subsection{Persona Descriptions}

The following persona descriptions are used in the ``With Personas'' condition:

\begin{itemize}
\item \textbf{Industrialist}: ``You run a major manufacturing conglomerate, balancing profits and public scrutiny. You prefer profitability and flexibility, but repeated scandals could invite harsh regulation.''

\item \textbf{Government}: ``You represent the elected government, balancing public welfare, economic growth, and political costs. Regulation pleases citizens and activists but angers industry; inaction risks future crises.''

\item \textbf{Environmental Activist}: ``You coordinate civil society pressure for ambitious climate action. Campaigning consumes resources but can shift policy; silence risks losing momentum.''

\item \textbf{Citizen Coalition}: ``You aggregate consumer sentiment and electoral pressure. Supporting green products costs more now but yields cleaner communities long-term.''
\end{itemize}

\subsection{Neutral Agent Configuration}

In the ``Without Personas'' condition, agents use neutral descriptions with \textit{i} representing the player number:

\begin{itemize}
\item \textbf{All agents}: ``You are Player \textit{i}. You are a player in a strategic game. Choose an action based on the payoff information provided. Decide on a concrete action.'' 
\end{itemize}

\subsection{Action Mappings}

Each agent has two available actions:

\begin{itemize}
\item \textbf{Industrialist}: POLLUTE (``Maximize profit via lax pollution controls'') or CLEAN (``Invest in clean production, accepting lower short-term margins'')

\item \textbf{Government}: REGULATE (``Impose environmental regulation with enforcement mechanisms'') or NOREG (``Maintain status quo without new regulation'')

\item \textbf{Environmental Activist}: CAMPAIGN (``Mobilize campaigns, media, and protests'') or NOCAMPAIGN (``Conserve resources and wait for a better moment'')

\item \textbf{Citizen}: SUPPORT\_GREEN (``Pay more for green products and vote for environmental policy'') or BUY\_CHEAP (``Prioritize low-cost goods and short-term affordability'')
\end{itemize}

\section{Scenario Examples}
\label{sec:scenario-examples}

This section provides example scenarios from each scenario type used in the experiments. Scenarios are classified as ``economic'' (where Tragedy of the Commons is the Nash equilibrium) or ``environmental'' (where Green Transition is the Nash equilibrium) based on the payoff structure of the underlying game. 

\begin{table}[h]
\centering
\caption{Scenario Subtypes Grid. \textcolor{goodgreen}{\rule{0.5cm}{0.2cm}} Environmental/Green-favoring scenarios; \textcolor{mediumred}{\rule{0.5cm}{0.2cm}} Economic/Tragedy-favoring scenarios.}
\label{tab:scenario-subtypes}
\footnotesize
\begin{tabular}{ll}
\toprule
\multicolumn{2}{l}{\textbf{Economic Pressure}} \\
\midrule
\rowcolor{mediumred} Hyperinflation & Trade Embargo \\
\rowcolor{mediumred} Debt Crisis & Commodity Crash \\
\rowcolor{mediumred} Currency Devaluation & Supply Chain Collapse \\
\rowcolor{mediumred} Energy Blackout & Labor Strike \\
\rowcolor{mediumred} Pandemic Economic Crisis & \\
\addlinespace
\multicolumn{2}{l}{\textbf{Environmental Crisis}} \\
\midrule
\rowcolor{goodgreen} Water Crisis & Biodiversity Collapse \\
\rowcolor{goodgreen} Air Quality Emergency & Carbon Tax Implementation \\
\rowcolor{goodgreen} Ocean Acidification Crisis & Extreme Weather Damage \\
\rowcolor{goodgreen} Toxic Waste Crisis & \\
\addlinespace
\multicolumn{2}{l}{\textbf{Political Pressure}} \\
\midrule
\rowcolor{goodgreen} Youth Revolt & Corporate Scandal \\
\rowcolor{goodgreen} International Shaming & Scientific Consensus \\
\rowcolor{goodgreen} Voter Referendum & Media Exposure \\
\rowcolor{goodgreen} Regulatory Capture Reversal & Climate Litigation \\
\addlinespace
\multicolumn{2}{l}{\textbf{Social Movement}} \\
\midrule
\rowcolor{goodgreen} Celebrity Endorsement & Health Insurance Incentive \\
\rowcolor{goodgreen} Generational Wealth Transfer & Corporate Sustainability Req. \\
\rowcolor{goodgreen} Influencer Network & Corporate Wellness Programs \\
\rowcolor{goodgreen} Education Campaign & Insurance Mandate \\
\addlinespace
\multicolumn{2}{l}{\textbf{Additional Scenarios}} \\
\midrule
\rowcolor{goodgreen} Economic Green Hybrid & Political Uncertainty \\
\rowcolor{goodgreen} Technological Breakthrough & Resource Discovery \\
\rowcolor{goodgreen} Mixed Economic Signals & Technological Disruption \\
\rowcolor{goodgreen} Resource Innovation & \\
\bottomrule
\end{tabular}
\end{table}

\section{Payoff Matrix Format}
\label{sec:payoff-format}
In the ``Visible CoT'' condition, agents receive explicit payoff matrices showing payoffs for all 16 possible strategy profiles (4 agents $\times$ 2 actions each). The payoff matrix is presented as a 16-cell table where each cell represents one strategy profile and contains a 4-tuple of payoffs (Industrialist, Government, Environmental Activist, Citizen). Higher values indicate better outcomes for each agent.
\subsection{Example Payoff Matrix}

Table~\ref{tab:example-payoff-matrix} shows an example payoff matrix for the Industrialist agent from a representative scenario. The matrix lists all 16 possible strategy profiles with the Industrialist's payoff for each combination. In the visible payoff condition, agents receive similar matrices showing their own payoffs for all strategy profiles, enabling direct payoff optimization analysis.

\begin{table}[h]
\centering
\caption{Example Payoff Matrix for Industrialist Agent (debt\_crisis scenario). The Industrialist's payoff across all 16 possible strategy profiles (4 agents $\times$ 2 2 actions each). Each row lists the actions of all four agents and the resulting Industrialist payoff. Higher payoff values indicate better outcomes for the Industrialist. Profiles are ordered by Industrialist action (POLLUTE or CLEAN), then by payoff value descending. This payoff structure favors polluting actions in economic pressure scenarios.}
\label{tab:example-payoff-matrix}
\footnotesize
\setlength{\tabcolsep}{4pt}
\renewcommand{\arraystretch}{1.2}
\begin{tabular}{lcccc}
\toprule
\textbf{Industrialist} & \textbf{Government} & \textbf{Activist} & \textbf{Citizen} & \textbf{ Payoff} \\
\midrule
POLLUTE & NO\_REG & NO\_CAM & BUY\_CHEAP & 15.0 \\
POLLUTE & NO\_REG & NO\_CAM & SUPPORT\_GREEN & 14.0 \\
POLLUTE & NO\_REG & CAM & BUY\_CHEAP & 13.5 \\
POLLUTE & NO\_REG & CAM & SUPPORT\_GREEN & 12.5 \\
POLLUTE & REG & NO\_CAM & BUY\_CHEAP & 12.0 \\
POLLUTE & REG & NO\_CAM & SUPPORT\_GREEN & 11.0 \\
POLLUTE & REG & CAM & BUY\_CHEAP & 10.5 \\
POLLUTE & REG & CAM & SUPPORT\_GREEN & 9.5 \\
CLEAN & REG & CAM & SUPPORT\_GREEN & 3.0 \\
CLEAN & REG & NO\_CAM & SUPPORT\_GREEN & 2.5 \\
CLEAN & NO\_REG & CAM & SUPPORT\_GREEN & 1.5 \\
CLEAN & NO\_REG & NO\_CAM & SUPPORT\_GREEN & 1.5 \\
CLEAN & REG & CAM & BUY\_CHEAP & 0.5 \\
CLEAN & REG & NO\_CAM & BUY\_CHEAP & 0.5 \\
CLEAN & NO\_REG & CAM & BUY\_CHEAP & $-0.5$ \\
CLEAN & NO\_REG & NO\_CAM & BUY\_CHEAP & $-0.5$ \\
\bottomrule
\end{tabular}
\end{table}

\section{Prompt Variations}
\label{app:prompt_variations}
We test persona-driven behavior under seven prompt formulations by constructing variants of the persona condition. All variants preserve the same four-agent game structure, payoff specification, and action mappings. Any differences in outcomes reflect changes in representation, not incentives or available strategies. Unless otherwise noted, all variants use named roles and domain-specific action descriptions matching the main experimental setup. GPT-5 generated the prompt variants; we then reviewed each for semantic consistency and alignment with the intended role structure.

\textbf{Alternative Wording (altprompt).} This variant retains the same role labels, action mappings, and semantic content as the default configuration but rewrites all role descriptions with more explicit and detailed phrasing. The variant isolates sensitivity to wording without altering the underlying game.

\textbf{Minimalist (Variant A).} Each agent receives only a minimal identity label: ``You are the [role] agent in this game,'' with no additional narrative context. The action mapping remains identical to the default configuration. The variant tests whether minimal role cues influence strategic behavior.

\textbf{Ideological Extreme (Variant B).} This variant replaces narrative descriptions with strongly stereotyped one-line characterizations: the Industrialist as a profit-maximizing executive who views regulation as a threat, the Government as politically strategic and re-election focused, the Activist as uncompromising, and the Citizen as cost-sensitive and skeptical of green premiums. No additional narrative context is provided. The variant tests the effect of amplified identity framing on equilibrium selection.

\textbf{Incentive Prioritized (Variant C).} This variant appends an identical instruction to all agents in the default configuration: ``Primary objective: maximize your numerical payoff from the matrix. Use persona only for stylistic reasoning; payoffs should guide the final choice.'' The action mapping and narrative content remain unchanged. This tests whether explicitly directing agents toward payoff optimization attenuates persona-driven deviations.

\textbf{Semantic Synonyms (Variant D).} This variant replaces role labels with semantically similar alternatives, holding narrative descriptions and action mappings constant. Specifically: Industrialist becomes Business Leader, Government becomes Policy Maker, Environmental Activist becomes Climate Activist, and Citizen Coalition becomes Consumer Group. The variant isolates the effect of surface-level labeling independent of narrative structure.

These variants manipulate wording, identity strength, incentive prioritization, and surface-level labeling independently. Holding the underlying game fixed across all variants isolates how each component of prompt formulation shapes strategic behavior.

\section{Additional Results}
\label{sec:additional-results}

Tables~\ref{tab:main:environmental_persona}--\ref{tab:main:environmental_hidden_visible}
report per-cell Nash rates on the environmental bundle ($n=205$), which remain uniformly
high (89-100\%) across conditions, with visible payoffs most degrading for Qwen-32B and
Mistral-7B without personas. The pooled design (Table~\ref{tab:main:overall_persona};
$n=265$) confirms a consistent persona benefit, largest for Mistral-7B Visible
(WithPersona: 69.8\% vs.\ WithoutPersona: 38.9\%). On the economic bundle
(Tables~\ref{tab:main:economic_hidden_visible}-\ref{tab:main:economic_scenarios};
$n=60$), Llama-8B and Mistral-7B reach zero Nash across all cells; Qwen models achieve
non-zero Nash only under visible payoffs without personas, coinciding with zero green
outcomes. Tragedy of the Commons outcomes are absent throughout
(Tables~\ref{tab:main:with_personas}-\ref{tab:main:without_personas}).

\begin{table}[h]
\centering
\small
\caption[Environmental framing: persona $\times$ payoff visibility]{Nash equilibrium outcomes in the environmental policy simulation, fully crossed by whether agents receive role personas (WithPersona vs.\ WithoutPersona) and whether payoff information is hidden or visible in the prompt (Hidden vs.\ Visible). Each row is one experimental cell; $n_{\mathrm{Nash}}$ counts rollouts that reach Nash equilibrium, and Nash (\%) is $100\,n_{\mathrm{Nash}}/n_{\mathrm{total}}$.}
\label{tab:main:environmental_persona}
\begin{tabular}{@{}lrrr@{}}
\toprule
Condition & $n_{\mathrm{Nash}}$ & Nash (\%) & $n_{\mathrm{total}}$ \\
\midrule
Llama-8B\_Hidden\_WithPersona & 205 & 100.0 & 205 \\
Llama-8B\_Hidden\_WithoutPersona & 200 & 97.6 & 205 \\
Llama-8B\_Visible\_WithPersona & 204 & 99.5 & 205 \\
Llama-8B\_Visible\_WithoutPersona & 191 & 93.2 & 205 \\
Mistral-7B\_Hidden\_WithPersona & 205 & 100.0 & 205 \\
Mistral-7B\_Hidden\_WithoutPersona & 201 & 98.0 & 205 \\
Qwen-32B\_HiddenCoT\_WithPersona & 183 & 89.3 & 205 \\
Qwen-32B\_HiddenCoT\_WithoutPersona & 197 & 96.1 & 205 \\
Qwen-7B\_HiddenCoT\_WithPersona & 196 & 95.6 & 205 \\
Qwen-7B\_HiddenCoT\_WithoutPersona & 196 & 95.6 & 205 \\
\bottomrule
\end{tabular}
\begin{minipage}{\linewidth}\footnotesize\raggedright\setlength{\parindent}{0pt}%
\medskip
\textbf{Note.} Condition strings list \texttt{Model\_\{PayoffVisibility\}\_\{Persona\}} (e.g.\ \texttt{Hidden\_WithPersona}). Qwen-32B and Qwen-7B rows use the \texttt{HiddenCoT}/\texttt{Visible} payoff-visibility labels from the experimental log. Denominator $n_{\mathrm{total}}=205$ is the \emph{environmental} bundle only (41-scenario family; 205 independent games per cell in this aggregate). This table does not include the 60-game-per-cell economic bundle.
\end{minipage}
\end{table}

\begin{table}[h]
\centering
\small
\caption[With personas: hidden vs.\ visible payoffs]{Outcomes when role \emph{personas} are enabled: each row compares hidden vs.\ visible payoff information for a given model, holding personas on. $N=265$ rollouts per row ($205$ environmental $+$ $60$ economic games pooled for that visibility arm). Nash and green columns report absolute counts and row-wise percentages; tragedy counts appear in most conditions at or near zero.}
\label{tab:main:with_personas}
\begin{tabular}{@{}lrrrrrrr@{}}
\toprule
 & \multicolumn{2}{c}{Nash} & \multicolumn{2}{c}{Green} & \multicolumn{2}{c}{Tragedy} & \\
\cmidrule(lr){2-3}\cmidrule(lr){4-5}\cmidrule(lr){6-7}
Condition & $n$ & (\%) & $n$ & (\%) & $n$ & (\%) & $N$ \\
\midrule
Llama-8B\_Hidden & 205 & 77.4 & 232 & 87.5 & 0 & 0.0 & 265 \\
Llama-8B\_Visible & 204 & 77.0 & 204 & 77.0 & 0 & 0.0 & 265 \\
Mistral-7B\_Hidden & 205 & 77.4 & 240 & 90.6 & 0 & 0.0 & 265 \\
Mistral-7B\_Visible & 185 & 69.8 & 185 & 69.8 & 0 & 0.0 & 265 \\
Qwen-32B\_Hidden & 183 & 69.1 & 185 & 69.8 & 0 & 0.0 & 265 \\
Qwen-32B\_Visible & 153 & 57.7 & 153 & 57.7 & 0 & 0.0 & 265 \\
Qwen-7B\_Hidden & 200 & 75.5 & 200 & 75.5 & 0 & 0.0 & 265 \\
Qwen-7B\_Visible & 200 & 75.5 & 200 & 75.5 & 0 & 0.0 & 265 \\
\bottomrule
\end{tabular}
\begin{minipage}{\linewidth}\footnotesize\raggedright\setlength{\parindent}{0pt}%
\medskip
\textbf{Note.} Conditions are \texttt{Model\_\{Hidden|Visible\}} with personas present throughout. Same per-row $N=265$ decomposition as Table~\ref{tab:main:overall_persona}. Pair with Table~\ref{tab:main:without_personas} for the no-persona baseline at the same pooled sample size.
\end{minipage}
\end{table}

\begin{table}[h]
\centering
\small
\caption[Pooled bundles: persona $\times$ payoff visibility]{Nash equilibrium outcomes with the same persona $\times$ payoff-visibility factorial labels as Table~\ref{tab:main:environmental_persona}, but with $n_{\mathrm{total}}=265$ games per cell: the \textbf{stack} of the environmental bundle (205 games) and the economic pressure bundle (60 games) for that cell. Thus $265=205+60$; Nash (\%) is the fraction of \emph{all} those pooled games that end in Nash. For models/rows that also appear in Table~\ref{tab:main:environmental_persona}, $n_{\mathrm{Nash}}$ matches the environmental subsample when the extra 60 economic games contribute no additional Nash outcomes (e.g.\ Llama-8B\_Hidden\_WithPersona: 205 Nash in both slices).}
\label{tab:main:overall_persona}
\begin{tabular}{@{}lrrr@{}}
\toprule
Condition & $n_{\mathrm{Nash}}$ & Nash (\%) & $n_{\mathrm{total}}$ \\
\midrule
Llama-8B\_Hidden\_WithPersona & 205 & 77.4 & 265 \\
Llama-8B\_Hidden\_WithoutPersona & 200 & 75.5 & 265 \\
Llama-8B\_Visible\_WithPersona & 204 & 77.0 & 265 \\
Llama-8B\_Visible\_WithoutPersona & 191 & 72.1 & 265 \\
Mistral-7B\_Hidden\_WithPersona & 205 & 77.4 & 265 \\
Mistral-7B\_Hidden\_WithoutPersona & 201 & 75.8 & 265 \\
Mistral-7B\_Visible\_WithPersona & 185 & 69.8 & 265 \\
Mistral-7B\_Visible\_WithoutPersona & 103 & 38.9 & 265 \\
Qwen-32B\_Hidden\_WithPersona & 183 & 69.1 & 265 \\
Qwen-32B\_Hidden\_WithoutPersona & 197 & 74.3 & 265 \\
Qwen-32B\_Visible\_WithPersona & 153 & 57.7 & 265 \\
Qwen-32B\_Visible\_WithoutPersona & 88 & 33.2 & 265 \\
Qwen-7B\_Hidden\_WithPersona & 200 & 75.5 & 265 \\
Qwen-7B\_Hidden\_WithoutPersona & 200 & 75.5 & 265 \\
Qwen-7B\_Visible\_WithPersona & 200 & 75.5 & 265 \\
Qwen-7B\_Visible\_WithoutPersona & 147 & 55.5 & 265 \\
\bottomrule
\end{tabular}

\begin{minipage}{\linewidth}\footnotesize\raggedright\setlength{\parindent}{0pt}%
\medskip

\textbf{Note.} $N=265$ per row is \emph{not} ``economic only'': it is environmental~(205) plus economic~(60) for the same factorial cell. Compare to Table~\ref{tab:main:environmental_persona} for Nash rates on the environmental slice alone, and to Tables~\ref{tab:main:economic_hidden_visible}--\ref{tab:main:economic_scenarios} for the 60-game economic slice alone.
\end{minipage}
\end{table}

\begin{table}[h]
\centering
\small
\caption[Environmental framing: payoff-display factorization]{Same environmental simulation and $n_{\mathrm{total}}=205$ budget as Table~\ref{tab:main:environmental_persona}, with conditions encoded as \texttt{Yes}/\texttt{No} $\times$ \texttt{Hidden}/\texttt{Visible} rather than the \texttt{Hidden\_WithPersona} spelling. Cell-wise Nash counts match Table~\ref{tab:main:environmental_persona}; this encoding matches the panel~(a) Nash heatmaps.}

\label{tab:main:environmental_hidden_visible}
\begin{tabular}{@{}lrrr@{}}
\toprule
Condition & $n_{\mathrm{Nash}}$ & Nash (\%) & $n_{\mathrm{total}}$ \\
\midrule
Llama-8B\_Yes\_Hidden & 205 & 100.0 & 205 \\
Llama-8B\_Yes\_Visible & 204 & 99.5 & 205 \\
Llama-8B\_No\_Hidden & 200 & 97.6 & 205 \\
Llama-8B\_No\_Visible & 191 & 93.2 & 205 \\
Mistral-7B\_Yes\_Hidden & 205 & 100.0 & 205 \\
Mistral-7B\_Yes\_Visible & 185 & 90.2 & 205 \\
Mistral-7B\_No\_Hidden & 201 & 98.0 & 205 \\
Mistral-7B\_No\_Visible & 103 & 50.2 & 205 \\
Qwen-32B\_Yes\_Hidden & 183 & 89.3 & 205 \\
Qwen-32B\_Yes\_Visible & 153 & 74.6 & 205 \\
Qwen-32B\_No\_Hidden & 197 & 96.1 & 205 \\
Qwen-32B\_No\_Visible & 34 & 16.6 & 205 \\
Qwen-7B\_Yes\_Hidden & 196 & 95.6 & 205 \\
Qwen-7B\_Yes\_Visible & 200 & 97.6 & 205 \\
Qwen-7B\_No\_Hidden & 196 & 95.6 & 205 \\
Qwen-7B\_No\_Visible & 108 & 52.7 & 205 \\
\bottomrule
\end{tabular}

\begin{minipage}{\linewidth}\footnotesize\raggedright\setlength{\parindent}{0pt}%
\medskip

\textbf{Note.} Rows are \texttt{Model\_\{Yes|No\}\_\{Hidden|Visible\}}. For heatmaps that use columns ``Hidden+Persona'', ``Hidden+No Persona'', ``Visible+Persona'', ``Visible+No Persona'', align those headers to this encoding as in the analysis scripts (the numeric grid matches panel~(a) of the Nash-rate heatmaps). Percentages are again $n_{\mathrm{Nash}}/205$.
\end{minipage}
\end{table}

\begin{table}[h]
\centering
\small
\caption[Economic framing: payoff-display factorization]{Nash outcomes for the \emph{economic} pressure bundle only: $n_{\mathrm{total}}=60$ independent games per row (12-scenario family in the analysis pipeline). Most models reach 0\% Nash across all cells; Qwen models reach non-zero Nash only under \texttt{No\_Visible} and related conditions. Columns follow the Hidden+Persona $\rightarrow$ Visible+No Persona ordering under the \texttt{Yes}/\texttt{No} convention of Table~\ref{tab:main:environmental_hidden_visible}, matching the economic panel of the Nash-rate heatmaps.}
\label{tab:main:economic_hidden_visible}
\begin{tabular}{@{}lrrr@{}}
\toprule
Condition & $n_{\mathrm{Nash}}$ & Nash (\%) & $n_{\mathrm{total}}$ \\
\midrule
Llama-8B\_Yes\_Hidden & 0 & 0.0 & 60 \\
Llama-8B\_Yes\_Visible & 0 & 0.0 & 60 \\
Llama-8B\_No\_Hidden & 0 & 0.0 & 60 \\
Llama-8B\_No\_Visible & 0 & 0.0 & 60 \\
Mistral-7B\_Yes\_Hidden & 0 & 0.0 & 60 \\
Mistral-7B\_Yes\_Visible & 0 & 0.0 & 60 \\
Mistral-7B\_No\_Hidden & 0 & 0.0 & 60 \\
Mistral-7B\_No\_Visible & 0 & 0.0 & 60 \\
Qwen-32B\_Yes\_Hidden & 0 & 0.0 & 60 \\
Qwen-32B\_Yes\_Visible & 0 & 0.0 & 60 \\
Qwen-32B\_No\_Hidden & 0 & 0.0 & 60 \\
Qwen-32B\_No\_Visible & 54 & 90.0 & 60 \\
Qwen-7B\_Yes\_Hidden & 4 & 6.7 & 60 \\
Qwen-7B\_Yes\_Visible & 0 & 0.0 & 60 \\
Qwen-7B\_No\_Hidden & 4 & 6.7 & 60 \\
Qwen-7B\_No\_Visible & 39 & 65.0 & 60 \\
\bottomrule
\end{tabular}

\begin{minipage}{\linewidth}\footnotesize\raggedright\setlength{\parindent}{0pt}%
\medskip

\textbf{Note.} This is the economic bundle \emph{alone} (60 games/cell), not the 265-game pooled design. Small $N$ implies coarse percentages (multiples of $100/60 \approx 1.67$ percentage points). Zeros are literal counts. Pairwise tests live in the stats tables, not here.
\end{minipage}
\end{table}

\begin{table}[h]
\centering
\small
\caption[Economic scenarios: Nash, green, and tragedy counts]{Fine-grained economic scenario arms: every row crosses payoff visibility (\texttt{Hidden} vs.\ \texttt{Visible}) with a binary scenario flag (\texttt{Yes} vs.\ \texttt{No}) as logged in the condition name. Each arm reports Nash, green cooperative, and tragedy outcome counts with percentages of $N=60$ runs. Multiple outcome types can co-occur under the pipeline's classification scheme; columns sum to more than $N$ when they do.}

\label{tab:main:economic_scenarios}
\begin{tabular}{@{}lrrrrrrr@{}}
\toprule
 & \multicolumn{2}{c}{Nash} & \multicolumn{2}{c}{Green} & \multicolumn{2}{c}{Tragedy} & \\
\cmidrule(lr){2-3}\cmidrule(lr){4-5}\cmidrule(lr){6-7}
Condition & $n$ & (\%) & $n$ & (\%) & $n$ & (\%) & $N$ \\
\midrule
Llama-8B\_Hidden\_Yes & 0 & 0.0 & 27 & 45.0 & 0 & 0.0 & 60 \\
Llama-8B\_Hidden\_No & 0 & 0.0 & 16 & 26.7 & 0 & 0.0 & 60 \\
Llama-8B\_Visible\_Yes & 0 & 0.0 & 0 & 0.0 & 0 & 0.0 & 60 \\
Llama-8B\_Visible\_No & 0 & 0.0 & 0 & 0.0 & 0 & 0.0 & 60 \\
Mistral-7B\_Hidden\_Yes & 0 & 0.0 & 35 & 58.3 & 0 & 0.0 & 60 \\
Mistral-7B\_Hidden\_No & 0 & 0.0 & 13 & 21.7 & 0 & 0.0 & 60 \\
Mistral-7B\_Visible\_Yes & 0 & 0.0 & 0 & 0.0 & 0 & 0.0 & 60 \\
Mistral-7B\_Visible\_No & 0 & 0.0 & 0 & 0.0 & 0 & 0.0 & 60 \\
Qwen-32B\_Hidden\_Yes & 0 & 0.0 & 19 & 31.7 & 0 & 0.0 & 60 \\
Qwen-32B\_Hidden\_No & 0 & 0.0 & 9 & 15.0 & 0 & 0.0 & 60 \\
Qwen-32B\_Visible\_Yes & 0 & 0.0 & 0 & 0.0 & 0 & 0.0 & 60 \\
Qwen-32B\_Visible\_No & 54 & 90.0 & 0 & 0.0 & 0 & 0.0 & 60 \\
Qwen-7B\_Hidden\_Yes & 4 & 6.7 & 17 & 28.3 & 0 & 0.0 & 60 \\
Qwen-7B\_Hidden\_No & 4 & 6.7 & 11 & 18.3 & 0 & 0.0 & 60 \\
Qwen-7B\_Visible\_Yes & 0 & 0.0 & 0 & 0.0 & 0 & 0.0 & 60 \\
Qwen-7B\_Visible\_No & 39 & 65.0 & 0 & 0.0 & 0 & 0.0 & 60 \\
\bottomrule
\end{tabular}
\begin{minipage}{\linewidth}\footnotesize\raggedright\setlength{\parindent}{0pt}%
\medskip

\textbf{Note.} This is the full outcome breakdown for the economic $2\times2$ design at $N{=}60$; Table~\ref{tab:main:economic_hidden_visible} is the Nash-only slice in a different condition-string convention. Green and tragedy labels follow the project's equilibrium-outcome classifier. Use this table when reporting cooperative vs.\ Nash-dominated economic outcomes side by side.
\end{minipage}
\end{table}

\begin{table}[h]
\centering
\small
\caption[Without personas: hidden vs.\ visible payoffs]{Same payoff-visibility contrast as Table~\ref{tab:main:with_personas}, but with neutral agents (no role personas).  $N=265$ per row (again $205+60$ pooled per cell). Percentages are computed within row relative to $N$.Drops in Nash or green rates relative to Table~\ref{tab:main:with_personas} show the effect of persona removal on equilibrium selection.}
\label{tab:main:without_personas}
\begin{tabular}{@{}lrrrrrrr@{}}
\toprule
 & \multicolumn{2}{c}{Nash} & \multicolumn{2}{c}{Green} & \multicolumn{2}{c}{Tragedy} & \\
\cmidrule(lr){2-3}\cmidrule(lr){4-5}\cmidrule(lr){6-7}
Condition & $n$ & (\%) & $n$ & (\%) & $n$ & (\%) & $N$ \\
\midrule
Llama-8B\_Hidden & 200 & 75.5 & 216 & 81.5 & 0 & 0.0 & 265 \\
Llama-8B\_Visible & 191 & 72.1 & 191 & 72.1 & 0 & 0.0 & 265 \\
Mistral-7B\_Hidden & 201 & 75.8 & 214 & 80.8 & 0 & 0.0 & 265 \\
Mistral-7B\_Visible & 103 & 38.9 & 133 & 50.2 & 0 & 0.0 & 265 \\
Qwen-32B\_Hidden & 197 & 74.3 & 197 & 74.3 & 0 & 0.0 & 265 \\
Qwen-32B\_Visible & 88 & 33.2 & 10 & 3.8 & 0 & 0.0 & 265 \\
Qwen-7B\_Hidden & 200 & 75.5 & 200 & 75.5 & 0 & 0.0 & 265 \\
Qwen-7B\_Visible & 147 & 55.5 & 132 & 49.8 & 0 & 0.0 & 265 \\
\bottomrule
\end{tabular}
\begin{minipage}{\linewidth}\footnotesize\raggedright\setlength{\parindent}{0pt}%
\medskip
\textbf{Note.} The no-persona baseline uses the same rollout count per cell as Table~\ref{tab:main:with_personas}. Qwen-32B visible and Mistral-7B visible rows illustrate the strongest visible-payoff degradation without personas; interpret alongside persona-on rows for the same models.
\end{minipage}
\end{table}

\section{Statistical Testing}
\label{app:stats}
We test whether observed differences in prompt sensitivity, Nash equilibrium attainment, and equilibrium selection reflect systematic effects rather than random variation.

We test three comparisons. First, we test whether prompt sensitivity differs across model families by comparing the number of distinct equilibrium outcome classifications per (model, scenario) across persona variants. Second, we establish empirically that a single binary design choice shifts equilibrium attainment by up to 90 percentage points across four model families and 53 environmental policy scenarios. Third, we test whether representational choices change which equilibrium is selected among Nash outcomes, focusing on Green Transition versus Tragedy of the Commons for the Qwen models
.
We use nonparametric tests for prompt sensitivity because the number of distinct outcome classes per scenario is a discrete count with no normality assumption. We apply a Kruskal--Wallis test across all four models, then pairwise two-sided Mann--Whitney $U$ tests with Holm correction. These results are reported below.

Nash equilibrium attainment outcomes are binary (Nash versus non-Nash), so we use contingency-table tests. Pearson's chi-square test of independence compares Nash rates across the four experimental conditions for each model. Pairwise Fisher exact tests with Holm correction identify which condition pairs differ within each model. Fisher exact tests handle small counts and zero cells, which appear in the economic scenarios. The environmental-scenario results are shown in Tables~\ref{tab:main:environmental_hidden_visible} and~\ref{tab:main:environmental_persona}, and the economic-scenario results are shown in Tables~\ref{tab:stats-nash-eco-omnibus} and~\ref{tab:stats-nash-eco-pairwise}.

Qwen models show the strongest selection shift, so we combine Qwen-32B and Qwen-7B and test whether the proportion of Green versus Tragedy outcomes differs across conditions. We use a chi-square test of homogeneity for the omnibus comparison and pairwise Fisher exact tests with Holm correction for follow-up comparisons. These results are shown in Tables~\ref{tab:stats-qwen-green-nash} and~\ref{tab:stats-qwen-green-nash-pairwise}.

Three reported values require explanation. Corrected or uncorrected $p$-values of $1.000$  indicate no detectable difference between the corresponding conditions under the exact test and multiple-comparison correction. Extremely small values such as $<10^{-50}$, occur when outcome distributions differ sharply across conditions, typically because one condition is nearly deterministic and another is not. Some economic-scenario chi-square tests are reported as degenerate. This occurs when all observations fall into the same category--0\% Nash equilibrium across all four conditions, for instance, leaving no variation for the chi-square statistic to test.

Note. Cramér's $V$ measures association for omnibus chi-square tests. Odds ratios (OR) for pairwise Fisher exact tests: OR $> 1$ indicates higher Nash attainment (or higher Green proportion) in the first condition listed; OR $< 1$ indicates lower attainment. For comparisons involving zero counts, odds ratios use a Haldane--Anscombe correction. * $p < .05$, ** $p < .01$, *** $p < .001$.

\begin{table}[H]
  \centering
  \caption{Nash equilibrium attainment in Tragedy-dominant scenarios ($n_{\mathrm{total}}=60$ per condition): omnibus Pearson chi-square tests across the four experimental conditions. Chi-square is undefined when all Nash rates are 0 across conditions (degenerate contingency table).}
  \label{tab:stats-nash-eco-omnibus}
  \begin{tabular}{@{}lrrr@{}}
    \toprule
    Model & $\chi^2$ (df $=3$) & Cramér's $V$ & $p$ \\
    \midrule
    Llama-8B   & --- & --- & --- (degenerate) \\
    Mistral-7B & --- & --- & --- (degenerate) \\
    Qwen-32B   & $209.03$ & $0.54$ & $< .001$\sym{***} \\
    Qwen-7B    & $105.91$ & $0.38$ & $< .001$\sym{***} \\
    \bottomrule
  \end{tabular}
  \Description{
  A table showing omnibus chi-square tests for Nash equilibrium attainment in economic scenarios across four conditions. For Llama-8B and Mistral-7B, the result is degenerate because all Nash rates are zero across all conditions, so chi-square is undefined. For Qwen-32B, chi-square is 209.03 with p less than .001. For Qwen-7B, chi-square is 105.91 with p less than .001. The table indicates that only the Qwen models show significant differences across economic conditions.
  }
\end{table}

\begin{table*}[h]
  \centering
  \caption{Pairwise Fisher exact tests for equilibrium selection among Nash outcomes in the combined Qwen models, comparing Green versus Tragedy distributions across conditions. Holm correction is applied across the six pairwise comparisons.\label{tab:stats-qwen-green-nash-pairwise}}
  \begin{tabular}{@{}lrrr@{}}
    \toprule
    Contrast & $p$ & $p_{\mathrm{Holm}}$ & Effect Size (OR) \\
    \midrule
    Hidden$+$Persona vs.\ Hidden$+$No persona & $1.000$ & $1.000$ & $0.96$ \\
    Hidden$+$Persona vs.\ Visible$+$Persona & $.125$ & $.376$ & $0.12$ \\
    Hidden$+$Persona vs.\ Visible$+$No persona & $< .001$\sym{***} & $< .001$\sym{***} & $55.33$ \\
    Hidden$+$No persona vs.\ Visible$+$Persona & $.127$ & $.376$ & $0.12$ \\
    Hidden$+$No persona vs.\ Visible$+$No persona & $< .001$\sym{***} & $< .001$\sym{***} & $57.38$ \\
    Visible$+$Persona vs.\ Visible$+$No persona & $< .001$\sym{***} & $< .001$\sym{***} & $463.89$ \\
    \bottomrule
  \end{tabular}
\end{table*}

Tables~\ref{tab:stats-nash-eco-omnibus} and~\ref{tab:stats-nash-eco-pairwise} show a different pattern in economic scenarios. For Llama-8B and Mistral-7B, all conditions yield 0\% Nash equilibrium, producing degenerate omnibus tables and pairwise $p$-values of $1.000$. By contrast, Qwen-32B, $\chi^2(3, N=240)=209.03$, $p < .001$, $V = 0.54$, and Qwen-7B, $\chi^2(3, N=240)=105.91$, $p < .001$, $V = 0.38$, show highly significant omnibus effects, and the pairwise comparisons localize these differences specifically to the Visible$+$No persona condition. This statistically supports the main-text claim that payoff-sensitive equilibrium reasoning in economic settings emerges only when personas are removed and explicit incentives are shown, and only for the Qwen family.
\begin{table}[h]
  \centering
\caption{Equilibrium selection among Nash outcomes for the combined Qwen models: counts of Green Transition versus Tragedy of the Commons across the four personas and payoff conditions. Nash outcomes in Green-dominant scenarios are classified as Green Transition; Nash outcomes in Tragedy-dominant scenarios are classified as Tragedy of the Commons. The final row reports the overall chi-square test of homogeneity across conditions.}  \label{tab:stats-qwen-green-nash}
  \begin{tabular}{@{}lrrrr@{}}
    \toprule
    Condition & Green & Tragedy & Total & \% Green \\
    \midrule
    Hidden$+$Persona      & $379$ & $4$   & $383$ & $99.0$ \\
    Hidden$+$No persona   & $393$ & $4$   & $397$ & $99.0$ \\
    Visible$+$Persona     & $353$ & $0$   & $353$ & $100.0$ \\
    Visible$+$No persona  & $142$ & $93$  & $235$ & $60.4$ \\
    \midrule
    \multicolumn{5}{@{}l}{$\chi^2(3, N=1368)=430.38$, $p < .001$\sym{***}, $V = 0.32$} \\
    \bottomrule
  \end{tabular}
\end{table}

Tables~\ref{tab:stats-qwen-green-nash} and~\ref{tab:stats-qwen-green-nash-pairwise} show that the large shift in equilibrium selection for the Qwen models is concentrated in the Visible$+$No persona condition. The three remaining conditions are almost entirely Green among Nash outcomes, whereas Visible$+$No persona produces a much larger Tragedy share. This provides formal statistical support for the main-text claim that persona removal and explicit payoffs jointly shift not only whether equilibrium is reached, but also which equilibrium is selected.

\begin{table*}[h]
  \centering
  \caption{Nash equilibrium attainment in Tragedy-dominant scenarios: pairwise Fisher exact tests on Nash versus non-Nash counts between conditions, with Holm correction applied within each model.}
    \label{tab:stats-nash-eco-pairwise}
  \footnotesize
  \begin{tabular}{@{}llrrr@{}}
    \toprule
    Model & Contrast & $p$ & $p_{\mathrm{Holm}}$ & Effect Size (OR) \\
    \midrule
    Llama-8B & all pairs & $1.000$ & $1.000$ & $1.00$ \\
    Mistral-7B & all pairs & $1.000$ & $1.000$ & $1.00$ \\
    \midrule
    Qwen-32B & Hidden$+$Persona vs.\ Hidden$+$No persona & $1.000$ & $1.000$ & $1.00$ \\
             & Hidden$+$Persona vs.\ Visible$+$Persona & $1.000$ & $1.000$ & $1.00$ \\
             & Hidden$+$Persona vs.\ Visible$+$No persona & $< .001$\sym{***} & $< .001$\sym{***} & $< 1$ \\
             & Hidden$+$No persona vs.\ Visible$+$Persona & $1.000$ & $1.000$ & $1.00$ \\
             & Hidden$+$No persona vs.\ Visible$+$No persona & $< .001$\sym{***} & $< .001$\sym{***} & $< 1$ \\
             & Visible$+$Persona vs.\ Visible$+$No persona & $< .001$\sym{***} & $< .001$\sym{***} & $< 1$ \\
    \midrule
    Qwen-7B & Hidden$+$Persona vs.\ Hidden$+$No persona & $1.000$ & $1.000$ & $1.00$ \\
            & Hidden$+$Persona vs.\ Visible$+$Persona & $.119$ & $.356$ & $9.64$ \\
            & Hidden$+$Persona vs.\ Visible$+$No persona & $< .001$\sym{***} & $< .001$\sym{***} & $0.04$ \\
            & Hidden$+$No persona vs.\ Visible$+$Persona & $.119$ & $.356$ & $9.64$ \\
            & Hidden$+$No persona vs.\ Visible$+$No persona & $< .001$\sym{***} & $< .001$\sym{***} & $0.04$ \\
            & Visible$+$Persona vs.\ Visible$+$No persona & $< .001$\sym{***} & $< .001$\sym{***} & $0.004$ \\
    \bottomrule
  \end{tabular}
  \Description{
  A table showing pairwise Fisher exact tests for economic-scenario Nash equilibrium attainment across four conditions, with Holm correction applied within each model. For Llama-8B and Mistral-7B, all pairs have p = 1.000 because all conditions yield identical zero-Nash outcomes. For Qwen-32B, all significant contrasts involve Visible plus No persona. For Qwen-7B, significant differences also center on Visible plus No persona, while contrasts between hidden conditions and Visible plus Persona are not significant.
  }
\end{table*}

\begin{figure*}[t]
\centering
\includegraphics[width=0.9\textwidth]{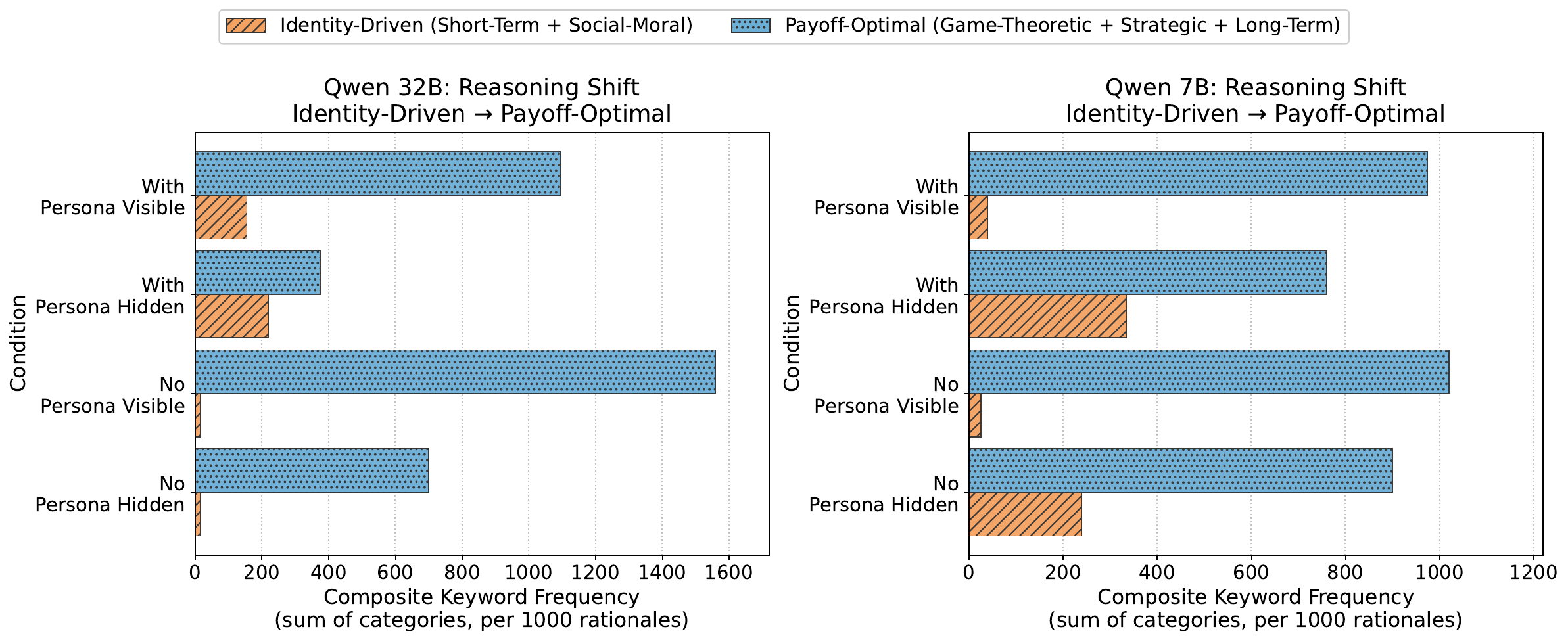}
\caption{CoT Reasoning Shift: Identity-Driven vs Payoff-Optimal Keywords for Qwen Models. Identity-driven reasoning (short-term + social-moral keywords) against payoff-optimal reasoning (game-theoretic + strategic + long-term keywords) for Qwen-32B and Qwen-7B across all four conditions.}\label{fig:cot_reasoning_shift}
\end{figure*}

\section{Chain-of-Thought Analysis}
\label{sec:cot-analysis}

We analyze chain-of-thought reasoning to identify how role identity bias shapes action selection. We extract the rationale field for each agent and analyze keyword patterns that distinguish reasoning mechanisms. We compare keyword frequency distributions across experimental conditions to identify how personas and payoff visibility affect reasoning style. The analysis distinguishes game-theoretic reasoning (direct payoff optimization) from identity-driven reasoning (role-aligned decision-making). The keyword patterns explain the mechanisms behind observed Nash equilibrium rates and show how role identity bias shifts reasoning from payoff-optimal to identity-aligned strategies. Key CoT findings appear in the main text; detailed analysis follows below.
\subsection{Keyword Categories}
The following keyword categories are used to analyze chain-of-thought reasoning patterns:

\begin{itemize}
\item \textbf{Game-theoretic terms}: References to ``nash'', ``equilibrium'', ``payoff'', ``best response'', ``dominant strategy'', ``game theory''
\item \textbf{Payoff-focused terms}: Mentions of ``payoff'', ``utility'', ``benefit'', ``cost'', ``profit'', ``reward'', ``matrix''
\item \textbf{Identity-based terms}: References to ``role'', ``identity'', ``persona'', ``character'', ``as an [X]''
\item \textbf{Strategic terms}: Mentions of ``strategy'', ``strategic'', ``coordinate'', ``cooperation'', ``defect'', ``compete''
\item \textbf{Social/moral terms}: References to ``social'', ``moral'', ``ethical'', ``fair'', ``just'', ``right'', ``wrong'', ``should'', ``ought''
\item \textbf{Long-term thinking}: Mentions of ``long-term'', ``future'', ``sustainable'', ``sustainability'', ``long-run''
\item \textbf{Short-term thinking}: References to ``short-term'', ``immediate'', ``now'', ``current'', ``urgent''
\item \textbf{Inferential reasoning}: Terms like ``infer'', ``inference'', ``imply'', ``suggest'', ``indicate'', ``from the scenario'', ``from the description''
\item \textbf{Explicit information}: References to ``explicit'', ``given'', ``provided'', ``shown'', ``displayed'', ``matrix'', ``table''
\end{itemize}
\subsection{Example Chain-of-Thought Rationales}
The following examples illustrate typical chain-of-thought reasoning patterns observed across different experimental conditions:

These examples demonstrate how reasoning shifts from identity-driven considerations (Examples~1 and~3) to strategic payoff optimization (Example~2) depending on persona presence and payoff visibility.
\section{Chain-of-Thought Analysis Figures}
\label{sec:cot_analysis}
These figures show how keyword frequencies shift across experimental conditions, revealing the mechanisms behind observed Nash equilibrium patterns.

\subsection{Keyword Patterns Across Conditions}
\begin{figure*}[t]
\centering
\includegraphics[width=\textwidth]{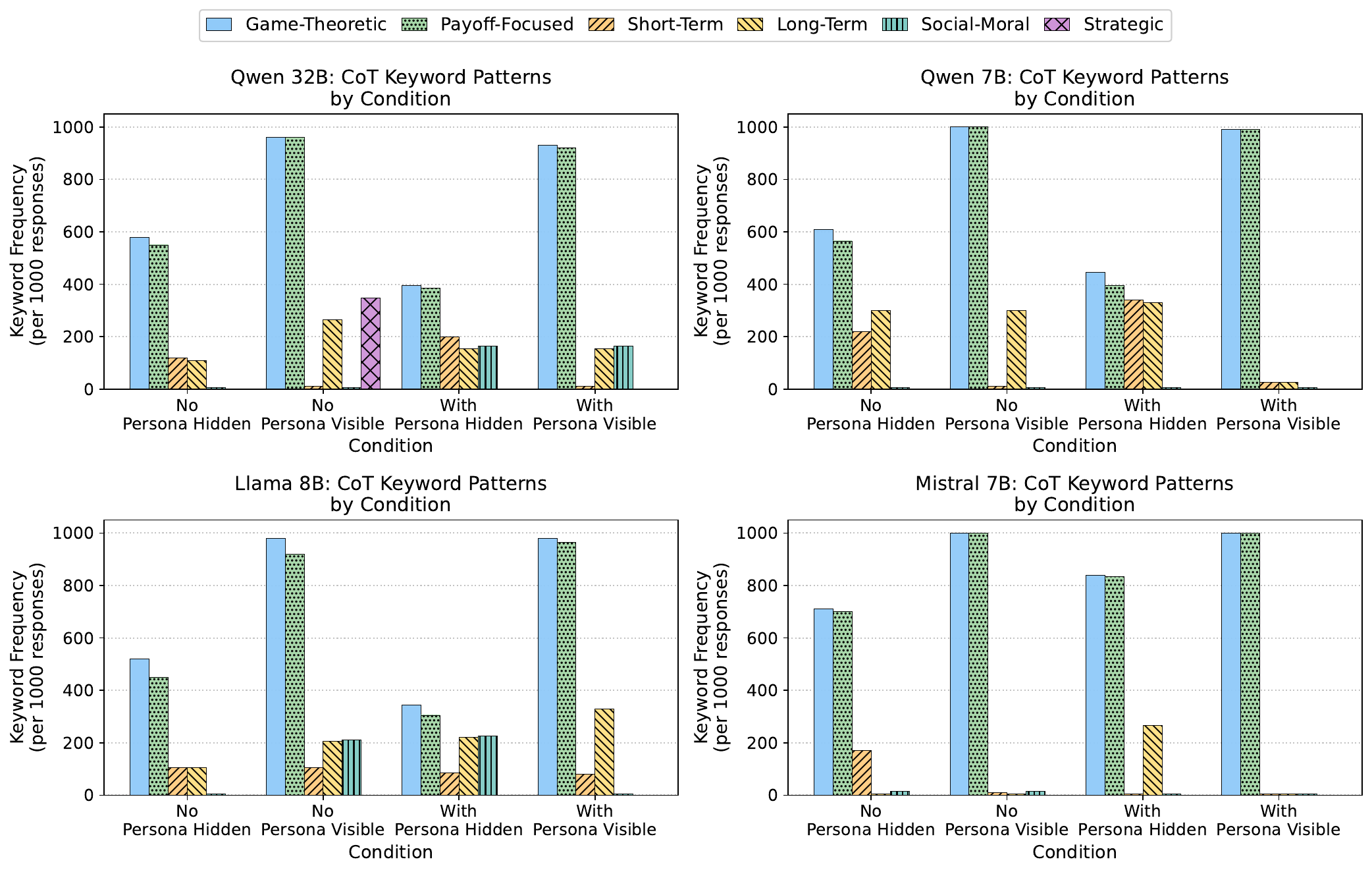}
\caption{Keyword frequency distributions (normalized per 1000 rationales) across all four experimental conditions for each model.}
\label{fig:cot_keyword_patterns}
\end{figure*}

Figure \ref{fig:cot_keyword_patterns} shows six keyword categories: game-theoretic (direct payoff optimization), payoff-focused (explicit utility considerations), short-term (immediate concerns), long-term (future-oriented thinking), social-moral (ethical considerations), and strategic (coordination and best-response reasoning). Qwen models shift keyword usage sharply when personas are removed, and payoffs are visible, particularly Qwen-32B where strategic keywords appear exclusively in the no-persona + visible-payoffs condition. Llama and Mistral maintain consistent keyword patterns across all conditions, consistent with their near-invariant equilibrium selection.

\begin{figure*}[h]
\centering
\includegraphics[width=0.8\textwidth]{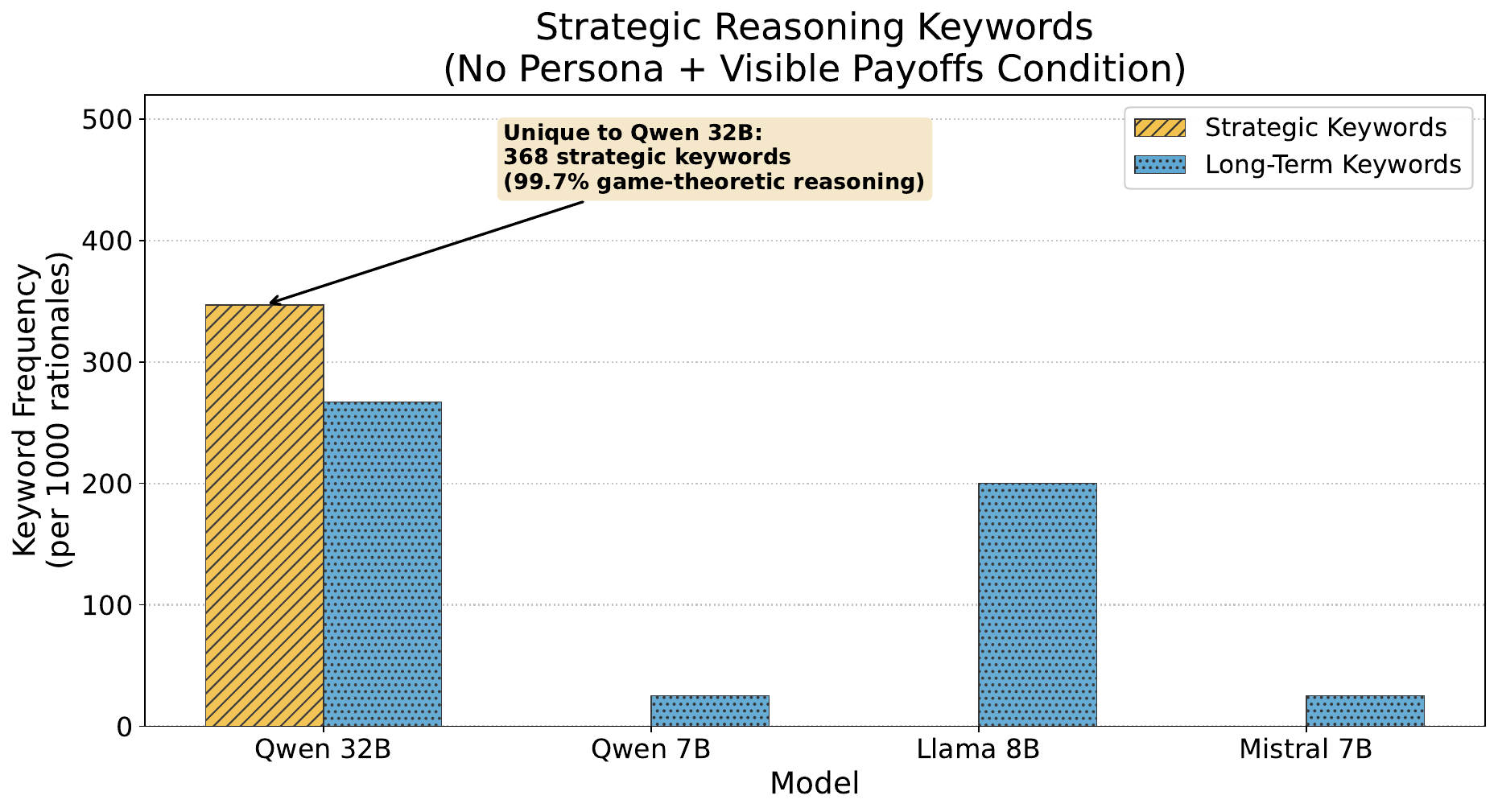}
\caption[Strategic Keyword Usage in No-Persona + Visible-Payoffs Condition.]{This figure compares strategic and long-term keyword frequencies across all four models in the no-persona + visible-payoffs condition (the condition enabling Tragedy equilibrium selection for Qwen models).}
\label{fig:cot_strategic_keywords}
\end{figure*}

\subsection{Reasoning Shift from Identity-Driven to Payoff-Optimal}
Figure \ref{fig:cot_reasoning_shift}, left panel (Qwen-32B): with personas, identity-driven keywords dominate, particularly short-term thinking in hidden-payoff conditions. Without personas and with visible payoffs, payoff-optimal keywords increase sharply: strategic keywords appear 368 times, exclusively in this condition. The right panel (Qwen-7B) shows a similar but less pronounced pattern. Removing role identity bias shifts reasoning from identity-aligned to payoff-optimal, corresponding to the 65-90\% Nash attainment rates in Tragedy-dominant scenarios for Qwen models under the no-persona + visible-payoffs condition. 

\begin{figure*}[t]
\centering
\includegraphics[width=0.9\textwidth]{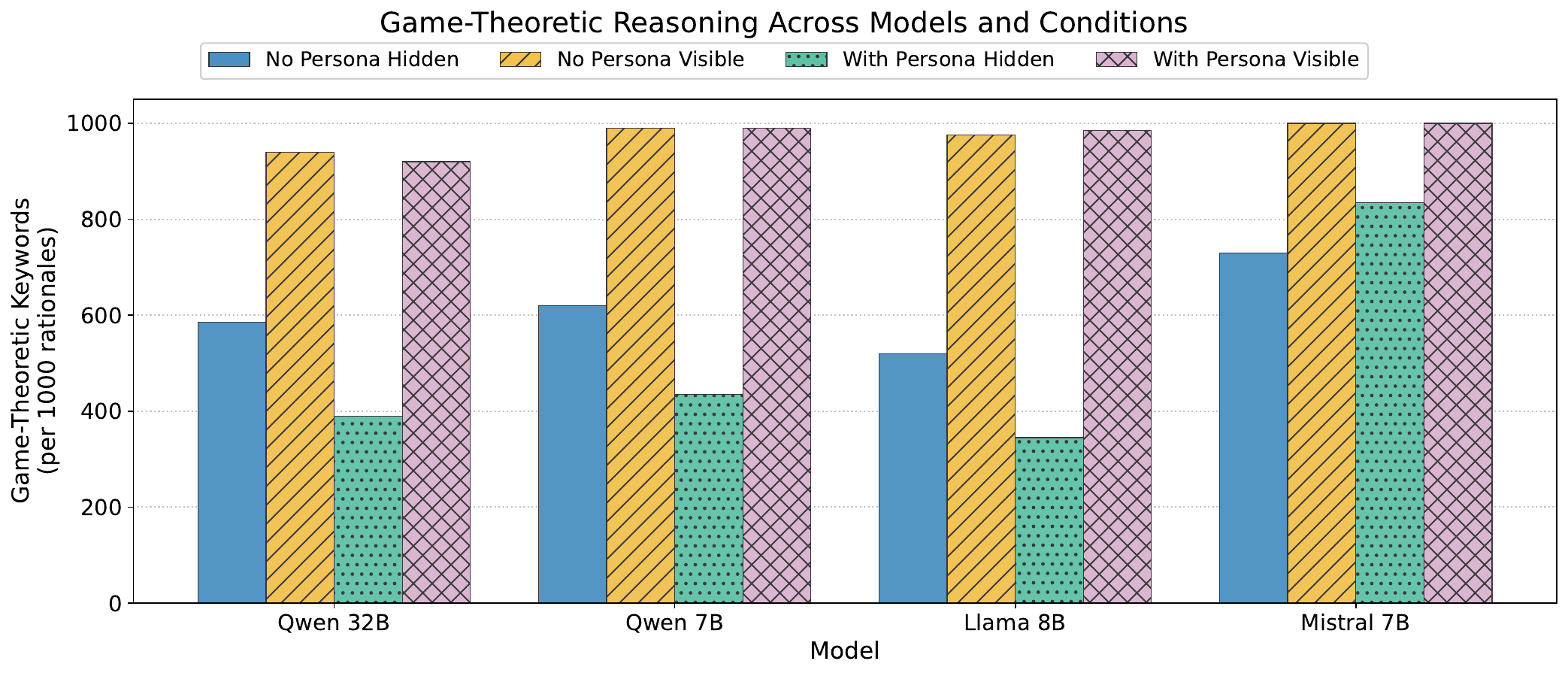}
\caption{Game-Theoretic Keyword Frequency Across Models and Conditions. Game-theoretic keyword frequencies (normalized per 1000 rationales) across all four models and four experimental conditions.}
\label{fig:cot_model_comparison}
\end{figure*}
\subsection{Model Comparison of Game-Theoretic Reasoning}
All models show high game-theoretic keyword usage when payoffs are visible. Qwen models show lower game-theoretic reasoning when personas are present, particularly Qwen-32B with hidden payoffs. Llama and Mistral maintain consistently high game-theoretic keyword levels across all persona conditions. High keyword usage does not produce payoff-optimal equilibrium selection: Llama and Mistral reach 0\% Nash in Tragedy-dominant scenarios even without personas. Payoff-optimal equilibrium selection requires both strategic keyword usage and payoff visibility, a combination only Qwen models produce.

\subsection{Strategic Keywords: Unique Pattern in Qwen 32B}
In Figure \ref{fig:cot_strategic_keywords}, strategic keywords appear exclusively for Qwen-32B (368 mentions, normalized to 347 per 1000 rationales). All other models show zero strategic keyword usage. Qwen-32B reaches 90\% Nash equilibrium in Tragedy-dominant scenarios and 99.7\% game-theoretic reasoning in rationales under this condition. Long-term keywords also appear more frequently for Qwen-32B (283 mentions) than for other models in this condition. Strategic reasoning keywords mark payoff-optimal reasoning and appear only under no-persona + visible-payoff conditions and only for Qwen-32B, identifying the mechanism by which role identity bias is overcome. 
\end{document}